\documentclass[reprint, aps, amsmath, amssymb]{revtex4-2}

\usepackage{epsfig}
\usepackage{graphicx}   
\usepackage{dcolumn}    
\usepackage{bm}         
\usepackage{color}
\usepackage{subcaption}

\usepackage{lmodern} 
\usepackage[bookmarks, colorlinks=false]{hyperref}
\usepackage{mathtools}
\hypersetup{colorlinks=true, allcolors=blue}
\usepackage[english]{babel}

\usepackage{siunitx}

\usepackage{multirow}
\usepackage{rotating}
\usepackage{tabularx}

\begin{document}

\title{Mechanical Model for a Full Fusion Tokamak Enabled by Supercomputing}

\author{W. M. E. Ellis}
\email{William.Ellis@ukaea.uk}
\affiliation{UK Atomic Energy Authority, Culham Campus, Oxfordshire OX14 3DB, UK}

\author{L. Reali}
\email{Luca.Reali@ukaea.uk, {corresponding author}}
\affiliation{UK Atomic Energy Authority, Culham Campus, Oxfordshire OX14 3DB, UK}

\author{A. Davis}
\email{Andrew.Davis@ukaea.uk}
\affiliation{UK Atomic Energy Authority, Culham Campus, Oxfordshire OX14 3DB, UK}

\author{H. M. Brooks}
\email{Helen.Brooks@ukaea.uk}
\affiliation{UK Atomic Energy Authority, Culham Campus, Oxfordshire OX14 3DB, UK}

\author{I. Katramados}
\email{Ioannis.Katramados@ukaea.uk}
\affiliation{UK Atomic Energy Authority, Culham Campus, Oxfordshire OX14 3DB, UK}

\author{A. J. Thornton}
\email{Andrew.Thornton@ukaea.uk}
\affiliation{UK Atomic Energy Authority, Culham Campus, Oxfordshire OX14 3DB, UK}

\author{R. J. Akers}
\email{Rob.Akers@ukaea.uk}
\affiliation{UK Atomic Energy Authority, Culham Campus, Oxfordshire OX14 3DB, UK}

\author{S. L. Dudarev}
\email{Sergei.Dudarev@ukaea.uk}
\affiliation{UK Atomic Energy Authority, Culham Campus, Oxfordshire OX14 3DB, UK}

\begin{abstract}
Determining stress and strain in a component of a fusion power plant involves defining boundary conditions for the mechanical equilibrium equations, implying the availability of a full reactor model for defining those conditions. To address this fundamental challenge of reactor design, a finite element method (FEM) model for the Mega-Ampere Spherical Tokamak Upgrade (MAST-U) fusion tokamak, operating at the Culham Campus of UKAEA, has been developed and applied to assess mechanical deformations, strain, and stress in the full tokamak structure, taken as a proxy for a fusion power plant. The model, handling 127 million finite elements using about 800 processors in parallel, illustrates the level of fidelity of structural simulations of a complex nuclear device made possible by the modern supercomputing systems. The model predicts gravitational and atmospheric pressure-induced deformations in broad agreement with observations, and enables computing the spectrum of acoustic vibrations of a tokamak, arising from mechanical disturbances like an earthquake or a plasma disruption. We introduce the notion of the {\it density of stress} to characterise the distribution of stress in the entire tokamak structure, and to predict the magnitude and locations of stress concentrations. The model enables defining computational requirements for simulating a whole operating fusion power plant, and provides a digital foundation for the assessment of reactor performance as well as for specifying the relevant materials testing programme. 
\end{abstract}

\maketitle

\onecolumngrid    

\section{Introduction}
International nuclear fusion development programme has reached a stage where a number of demonstration or even net power-generating fusion devices are expected to be commissioned over the next decade \cite{Meschini2023,Kim2020,Zheng2022,Mailloux2022,Chapman2024}. Extensive effort involving the use of the available tokamak facilities to support the ITER programme \cite{Mailloux2022,Joffrin2024}, materials and component-scale tests for generating the engineering design data \cite{Pintsuk2022,Cappelli2023,Federici2023,Waldon2024}, and the identification of critical aspects of a mature reactor design \cite{Hassanein2021} aim at establishing a regulatory framework for the commercial fusion power generation. 

Historically, the evolution of fusion devices has been driven by empirical scaling laws of plasma confinement, reflected in the heuristic construction of successive generations of fusion devices.  The notion of Integrated Tokamak modelling still refers to models for plasma dynamics \cite{Poli2018}, with structural engineering models exploring individual device components, for example the superconducting magnets \cite{Kim2020}, divertor targets \cite{You2021}, or breeding blankets \cite{Shimwell2019,Giancarli2020}. Reactor structure  optimisation, reflecting the operating conditions in the whole device and similar to the computational optimisation of electromagnetic coils in a fusion stellarator \cite{Kaptanoglu2024}, has not yet been explored. Notably, a component-scale view of a reactor not only makes it virtually impossible to pose the mechanical equilibrium problem, but it also makes it difficult to determine the operating conditions for which engineering tests need to be conducted. 

The macroscopic parameters specifying the operating conditions for materials are the temperature, stress, the radiation dose rate, the integral accumulated dose, the steady-state and time-dependent magnetic and gravitational loads, and the coupled chemical and radiation environments involving for example variable hydrogen isotope content in materials exposed to irradiation \cite{Mason2021}. These parameters vary depending on spatial location, and are self-consistently related through the operating conditions in a fusion device. It is not possible to accurately define the range of these parameters {\it a priori} from qualitative estimates based on the consideration of individual reactor components taken in isolation. The definition of representative local operating conditions for components in a power plant {\it and} the frequency of occurrence of these conditions in a multi-dimensional operating parameter space, is a natural foundation requirement for a fusion materials testing programme \cite{Knaster2016,Cappelli2023}. The extreme nature of operating conditions in a fusion power plant also implies a high rate of ageing of materials, an issue well recognised in nuclear fission  \cite{Odette2001,Spencer2021}, highlighting the fact that a sensible model for a fusion power plant must be an evolving entity, describing ageing of the structure and resulting changes in operating conditions over the reactor lifetime.               

Creating a predictive numerical representation of a fusion power plant, suitable for projecting its operation into the future over a realistic timescale, is the first step towards creating its digital twin \cite{Chiachio2022}. A virtual representation of a reactor requires capturing the operation of the {\it whole} device, and cannot be achieved by the more traditional methods whereby individual parts are abstracted away or where sub-modelling is deployed. A structural digital twin is fundamentally holistic; it involves modelling the structure of the entire object and predicting its response to variations of its environment, for example applied mechanical or magnetic loads, exemplified by plasma disruptions. The assessment of a response may involve optical or acoustic sensors relaying data back to the digital twin model in real time. The digital twin methodology has already been applied to large engineering structures \cite{Chiachio2022, Torzoni2024} and aerospace systems \cite{Glaessgen2012}. The latter example is particularly pertinent, as aerospace vehicles occasionally find themselves in an environment that was not explored in the design process, and that could affect materials in a way that was not previously assessed experimentally, for instance during long-duration space missions where there are no historic test data. Visual or other direct inspections may not be possible in some of the cases. This motivates high-fidelity simulations, similar to those driving the design of commercial fusion and fission power plants \cite{Meschini2023,Tian2022}, fusion testing facilities \cite{Tindall2023}, and the development of digital tools for controlling tokamak operation \cite{Kwon2022}.

A pioneering REacteur Virtuel d'Etudes (REVE) project was launched by EdF in France in the late 1990s \cite{Jumel2000,Jumel2002,Jumel2005} in response to concerns about the safety of operation of nuclear reactors beyond their projected lifetime, and about the decreasing availability of experimental facilities suitable for performing experimental tests on reactor materials. Numerical modelling of radiation effects was seen as a basis for reactor lifetime extension decisions \cite{Jumel2000}. A broadly similar dilemma is now encountered by commercial fusion, where the design of a fusion reactor requires not only the availability of extensive engineering data on how the properties of materials change during reactor operation, but also the specification of the operating conditions themselves. The Virtual Reactor concept now needs extending beyond the microstructural models for materials formulated and explored so far \cite{Jumel2000,Jumel2002,Jumel2005}, to enable a holistic treatment of the full reactor structure, required for defining the changing operating environments for materials and components over the expected lifetime of the device.         

A fusion reactor converts the kinetic energy of neutrons\textemdash \SI{14.1}{\mega\electronvolt} each, produced in a collision between a deuterium and tritium nuclei in the plasma \textemdash first into heat, generated when neutrons slow down and release their energy in the reactor materials, and subsequently into electricity. Hence, coupling neutron transport and structural thermo-mechanical simulations is a foundational step in any large-scale digital analysis of operation of the full device \cite{Brooks2022}. Since plasma confinement involves using strong magnetic fields, in addition to the intense heat loads and irradiation, magnetic forces act on the tokamak components \cite{Thome1982,Gruber1991,Bongiovi2018}. These stem from the steady-state currents, from the transient currents due to plasma disruptions, and from the ferromagnetic nature of steels used in radiation-resistant fusion blankets \cite{Mergia2008}. Other loads of mechanical origin include seismic accidents \cite{Ricciardi2009} or the transient pressure variations in the vacuum vessel, involving the loss of vacuum \cite{Bachmann2017}.

\begin{figure}[t]
  \includegraphics[width=0.99\columnwidth]{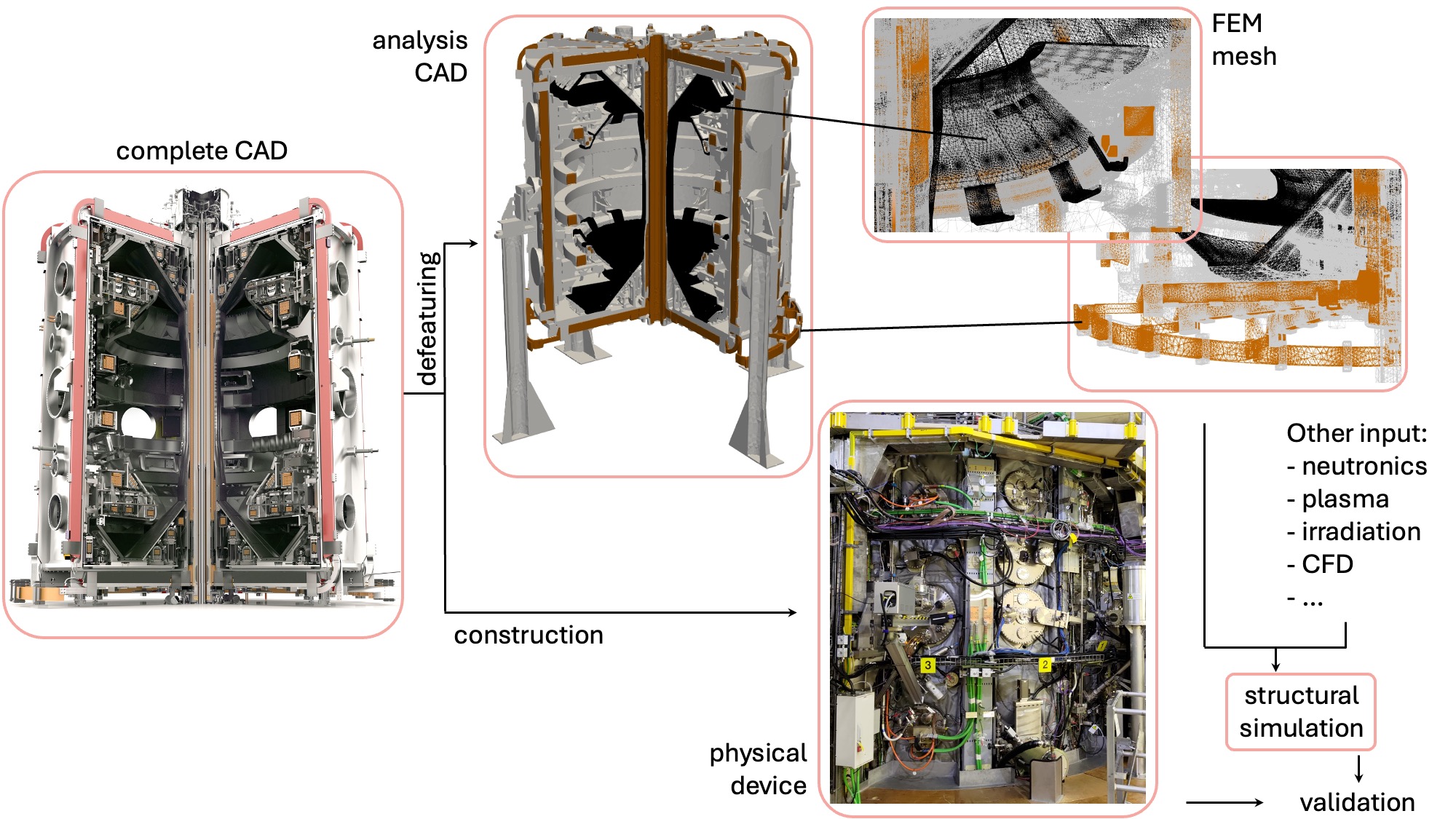} %
    \caption{Sketch illustrating a full tokamak simulation using the MAST-U spherical tokamak as a case study. A structural simulation requires the definition of loads as input, which in turn depend on the physical nature of the processes that a model intends to capture. In a tokamak, input might take the form of gravitational and magnetic forces, transient events of plasma-wall interaction, exposure to neutron irradiation, and variable thermal exposure. Given the multitude of input loading conditions present at the same time, the simulation process generates a mechanical digital twin,  which would then process information in real time from a real device via sensors for continuous validation, extrapolation and control.}
    \label{fig:twin_loop} 
\end{figure}

Figure \ref{fig:twin_loop} illustrates the logic of the study below, focused on the analysis of mechanical loads in a fusion tokamak. We start by taking a detailed CAD (computer-aided design) representation of the full tokamak device and de-featuring it, to make it tractable by the meshing software. The mesh is then taken by the finite element method (FEM) solver together with other input data such as material properties. Simulation results are validated using the available measurements taken on the physical device. Subsequently, the potential sources of uncertainties are established and analysed, contributing to the uncertainty quantification (UQ) of the model.

Here, we focus solely on the structural mechanics aspects of the tokamak model, on its consistency and connectivity as well as on the complexity of the device and resulting statistical aspects of mechanical loads. Effects of neutron exposure are treated in detail in a separate study. To provide an example of a realistic application of a full device simulation methodology, we model a real large-scale tokamak, namely the MAST-U tokamak operating at the Culham Campus of UKAEA. 

\section{Construction of the model}

The Mega-Ampere Spherical Tokamak (MAST) was operational between 1999-2013 \cite{Sykes2001,Katramados2011}. It ceased operations and was upgraded over the period from 2013 to 2020 \cite{Milnes2015}. MAST Upgrade (MAST-U) tokamak device is now fully operational \cite{Harrison2024}. The tokamak is constructed primarily of non-magnetic stainless steel, with copper magnets and carbon-tiled double divertor at the top and the bottom of the plasma chamber. The cylindrical chamber of the device is suspended on four supporting legs attached to a concrete foundation. The total weight of the device is estimated to be about 110 tons.

The structural complexity of a tokamak device, involving thousands of individual parts and contact surfaces, constitutes a new fundamental aspect of structural simulations in comparison with models for individual fusion reactor components \cite{You2021,Kim2020} or fission reactor pressure vessels \cite{Spencer2021,Tian2022}. The structural complexity leads to the FEM model becoming highly multiscale in relation to the varying size of the individual tetrahedral elements, resulting in a large overall number of elements that in the FEM representation shown in Figure \ref{fig:MASTU_sketch} is 127 million, equivalent to 83 million degrees of freedom (DOFs) for a purely mechanical calculation.   

The large size of the structural simulation model, requiring substantial computer power for performing the analysis, is not the only significant aspect illustrating its complexity. 
In a deuterium-tritium fuelled operating fusion power plant, materials at various parts of the structure are going to be evolving under different conditions, self-consistently determined by the model itself, through the thermal conductivity of reactor parts, the temperature dependence of permeation rates of hydrogen isotopes, the variation of the neutron exposure dose and neutron exposure dose rate changing as functions of spatial location, the strongly temperature dependent diffusion mobility of radiation defects, and the resulting strongly non-linear evolution of stress on the scale of the entire device. The constitutive laws of evolution of materials under such conditions are strongly non-linear, and the range of conditions themselves in the relevant parameter space is far more diverse in comparison even with the problem of structural evolution of zirconium cladding in a fission reactor \cite{Holt1988}, considered to be one of the greatest challenges in the field of fission. In the present study of a full tokamak structure, we only treat the linear elastic response of the materials, and  still the model offers an insight into what can be achieved by means of a large-scale computer simulation of a full highly complex multi-component fusion tokamak device. 

On the basis of the study below, we argue that virtual full device simulations offer a cost-effective and efficient option for assessing the performance and longevity of competing fusion reactor designs, reducing the waste associated with heuristic device construction, and offering a potentially faster route towards the deployment of fusion power generation.

\section{Finite element model for a full tokamak device}
The development of a satisfactory finite element model for a complex device remains an unfortunately manual process. The main steps include: the generation of CAD (Computer Aided Design) geometry, the identification and enforcement of contacts, and the discretisation of the imprinted geometry into a finite element mesh. Contacts are enforced by what is known as imprinting and merging the relevant CAD volumes \cite{White2004}. Imprinting forces adjacent surfaces to have same topology. The contact surfaces become a single, shared one by merging. The procedure results in a mesh that is conformal across volumes. In this context, a volume refers to a discrete constituent part of the CAD assembly. While the generation of a tetrahedral mesh is a largely automated process, preparation of a complex CAD model for imprinting and merging requires significant human effort.

\subsection{Model Preparation Process}
\subsubsection{CAD Model Preparation} 
The first step taken in developing the MAST-U finite element model was the preparation of a suitable geometrical CAD representation of the tokamak structure. An existing model previously used for neutron transport analysis was used as the basis for the MAST-U CAD representation. This model already contained numerous key components, including the vacuum vessel as well as much of the first wall shielding. However, many of the components included in the neutron transport model lacked mechanical support, and a significant number of components present in the real tokamak device were completely absent. These issues were rectified by creating CAD models of the missing components using reference drawings and JT models \cite{JTFiles} from the UKAEA Technical Document Management System (TDMS). For a more in depth description of the components added see Section \ref{ModelDescription}.

\subsubsection{Mesh preparation}
Coreform Cubit 2024.3 \cite{Cubit} was used to prepare the CAD for meshing and to generate the finite element mesh for the entire MAST-U device. Generating a finite element mesh from a CAD model is not equivalent to creating individual FEM representations of the individual components and hence forming the full structure. Following that procedure would have resulted in adjacent component meshes not conforming to one another. Appendix \ref{ImprintingAppendix}, Figure \ref{fig:OverlappingMeshes} gives an example of the undesirable outcome resulting from an attempt to mesh individual components taken in isolation. The unsatisfactory outcome manifests itself as the field values not crossing the component boundaries, which for mechanical simulations result in no mechanical contact between the components, leaving them to separate and fall if the structure is gravitationally loaded. Notably, the lack of correct mechanical contact behavior does not present an issue for a model designed solely for neutron transport simulations, and can only be identified by testing the model under a variety of mechanical loads.  

To generate a mesh with conformal elements across component boundaries, coincident component surfaces must be merged together. For merging to take place, surfaces must have like topology. Imprinting involves superimposing curves from surfaces on to neighboring co-planar surfaces in order to force like topology. Once like topology is ensured, these surfaces can then be merged together, which will result in a conformal mesh as shown in Appendix \ref{ImprintingAppendix}, Figure \ref{fig:Imprinting example}. Coreform Cubit allows users to perform imprinting and merging on imported CAD geometry, hence its use for the generation of the MAST-U finite element model was instrumental. Crucially, Cubit also allows for tolerant imprinting and merging. Tolerant imprinting allows for geometrical entities such as surfaces, curves and vertices to be imprinted on to neighboring geometry within a given distance, or tolerance. This is vital for the preparation of large CAD assemblies as tolerance issues within the CAD often result in small gaps and overlaps being present in the model, which would be problematic for standard imprinting.

Ideally, once all the necessary volumes are imprinted and merged, the geometry can be meshed. In practice, the process of generating a consistent FEM mesh for a complex multi-volume assembly similar to a tokamak structure, and consisting of hundreds of individual components, is cyclical. Errors encountered during meshing are often indicative of problems with the input CAD geometry, which a user must address and resolve before attempting to mesh again, at which point they may encounter more errors which they then must go back and rectify. One common cause of such errors are the incorrectly imprinted and merged surfaces. Other potential sources of error include small sliver surfaces or small curves present in the CAD model, which demand a relatively much smaller mesh sizing in comparison to the rest of the geometry. Components crossing CAD kernel boundaries between different applications can also become corrupted, resulting in failed meshing attempts, which are difficult to diagnose due a lack of visual anomalies prompting an investigation. When dealing with thousands of individual volumes, or parts, in a complex structure, this process can become immensely time consuming, with the time taken to prepare the model often vastly outweighing the time taken to run simulations on the final mesh \cite{IGA_2}. As the computational resources capable of running large scale finite element models become increasingly more accessible, the commonality of these large scale analyses will likely be limited by the current model preparation process, which is at best undesirable and at worst unacceptable.

\subsubsection{Mesh Generation}
After the volumes in the CAD geometry are correctly imprinted and merged, the model is ready to be meshed. Given the size and complexity of the MAST-U tokamak, we opt for a linear tetrahedral mesh. Tetrahedral meshes generally take significantly less time to generate compared to hexahedral meshes, especially when meshing complicated geometry that is not easily mapped \cite{Bernard2014}. One shortcoming of linear tetrahedral elements is their propensity to exhibit over-stiffening behaviour, referred to as volumetric locking. It should be noted that this behaviour is also exhibited by linear hexahedral elements, so it is not a limitation exclusive to linear tetrahedra. MOOSE implements B-bar correction to help account for this over-stiffening \cite{Hughes1987}. The mesh that we use in this study has been generated using Coreform Cubit and its implementation of TetMesh \cite{Cubit}.

\begin{figure}
    \centering
    \begin{subfigure}{0.49\linewidth}
        \includegraphics[width=\linewidth]{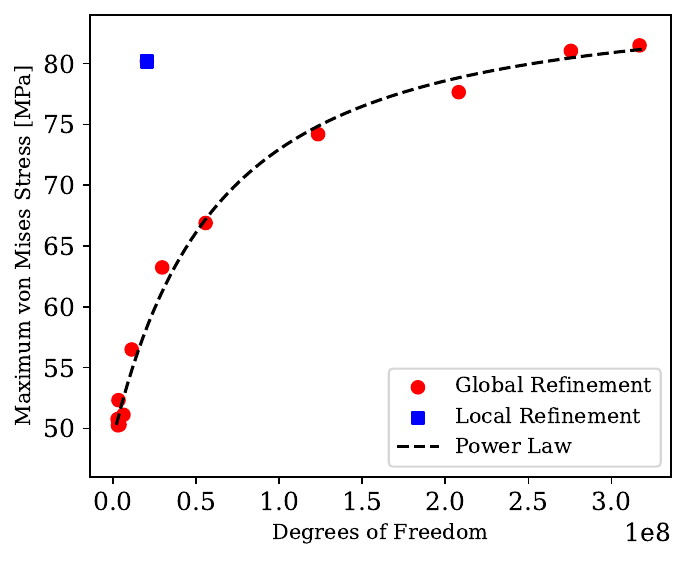}
        \caption{}
        \label{ConvVM}
    \end{subfigure}
    \begin{subfigure}{0.49\linewidth}
        \includegraphics[width=\linewidth]{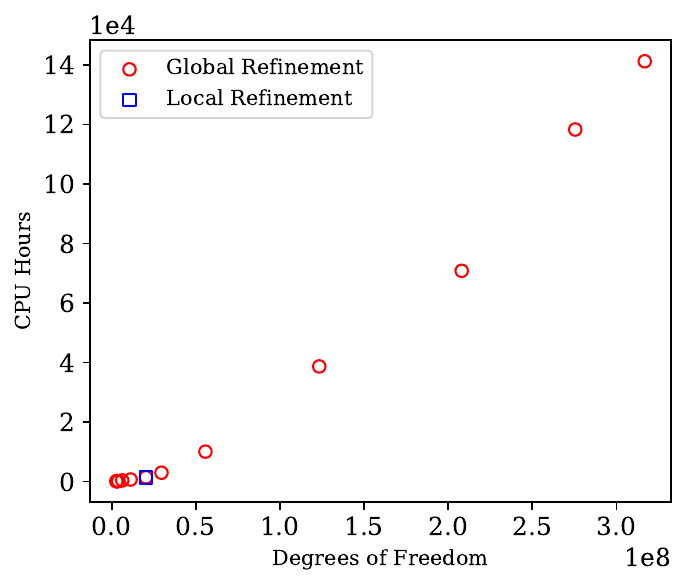}
        \caption{}
        \label{fig:ConvTime}
    \end{subfigure}
    \caption{Results from a convergence study conducted using a simplified MAST-U FEM model. The results show (a) the maximum von Mises stress predicted by the model versus the number of degrees of freedom in the model. The dashed line is a power law fit to the data, where the quality of fit remains nearly the same for the power law exponents varying in the interval from 1.5 and 2. (b) The total number of CPU hours required to arrive at a convergent result using a non-linear solver, versus the number of degrees of freedom in the corresponding FEM mesh for the structural model. In the limit of high number of degrees of freedom the scaling is broadly linear. The local refinement provides an optimal minimisation of computational cost without sacrificing accuracy.}
    \label{ConvergenceStudy}
\end{figure}

The MAST-U model developed here involves 3,938 individual volumes or, equivalently, small  components. In order to determine what sizing parameters to use when meshing a complex structure consisting of so many components, a mesh convergence study \cite{Melosh1990,Morin2002} was conducted. A simplified version of the entire MAST-U model with fewer volumes than the final model was used to carry out the convergence study. Using a simplified version of the model enabled a larger range of mesh resolutions to be tested, as meshing and simulation time was reduced compared to the fully featured model. This model was meshed using a variety of mesh sizing parameters to help identify the element size necessary to fully resolve the highest values of the von Mises stress predicted by the model when it is applied to mechanical equilibrium simulations. In the testing scenario the model is loaded by gravitational forces, with Dirichlet boundary conditions forcing zero displacements at the bottom surface of each supporting leg. Figure \ref{ConvergenceStudy} shows the results obtained from the mesh convergence study. The predicted maximum value of the von Mises stress computed using the simplified model is plotted against the number of degrees of freedom ${\sf N}$ characterising the FEM mesh. When using global refinement, the mesh convergence follows a power law of the form approximated by equation
\begin{equation}
    \sigma _{vM} ({\sf N})  =  \sigma _{vM}( \infty ) - {\alpha \over (\beta +{\sf N})^n} .
    \label{eq:PowerLawMesh}
\end{equation}
The left hand side of this equation represents the highest von Mises stress value predicted by the model, $\alpha$ and $\beta$ are constants and $n$ is the power law exponent that quantifies how quickly the stress value converges with increasing ${\sf N}$. In this case, choosing the power low exponent $n$ in the interval $[1.5,2]$ produces similar fits to the data. In Figure \ref{ConvVM} we use $n=1.50$, $\alpha=32.93$ and $\beta=0.94$. The point in blue refers to a model with a mesh that has received local mesh refinement in specific areas, for example areas of very high stress, in contrast to the red points that refer to models where the element size was modified globally.




As shown in Figure \ref{ConvergenceStudy}, the locally refined mesh predicts the higher values of the von Mises stress also predicted by the mesh with the most detailed global refinement, while requiring significantly fewer degrees of freedom (DOFs) to do so. DOFs in this context refer to the number of unknowns in the finite element equation system. Refining elements locally rather than globally allows us to capture the stress behaviour in specific areas of the model without drastically increasing the total number of DOFs, therefore reducing the computational requirements of the simulations as well as the time required to mesh the model. Figure \ref{fig:ConvTime} plots the number of CPU hours taken to run a displacement simulation against the number of DOFs in each mesh used in the convergence study. The number of CPU Hours here is calculated as
\begin{equation}
    \text{CPU Hours} = \text{MPI Tasks} * \text{Total Runtime}.
    \label{CPUHours}
\end{equation}
The mesh that had undergone a local refinement produces the convergent value of the von Mises stress similar to the most globally refined mesh, while requiring only 1.02\% of the computer resources to achieve convergence.

The results of the mesh convergence study demonstrate that refining the mesh does result in the maximum von Mises stress values exhibiting systematic convergence \cite{Melosh1990,Morin2002}. This firstly indicates that the highest values of stress in the model are being correctly resolved and are not caused by stress singularities. If stress singularities were present, the expected behaviour would be an exponential increase in stress as meshing resolution is increased. Secondly, the von Mises stress can be adequately evaluated via the local mesh refinement, reducing the time required to generate the mesh as well as the computational requirements of simulations when compared to a global mesh refinement. Given these results, the final mesh used for the analysis utilises the local element refinement to exploit the above advantages in numerical convergence.

\begin{figure}[h!]
    \centering
    \begin{subfigure}{0.49\linewidth}
        \includegraphics[width=\linewidth]{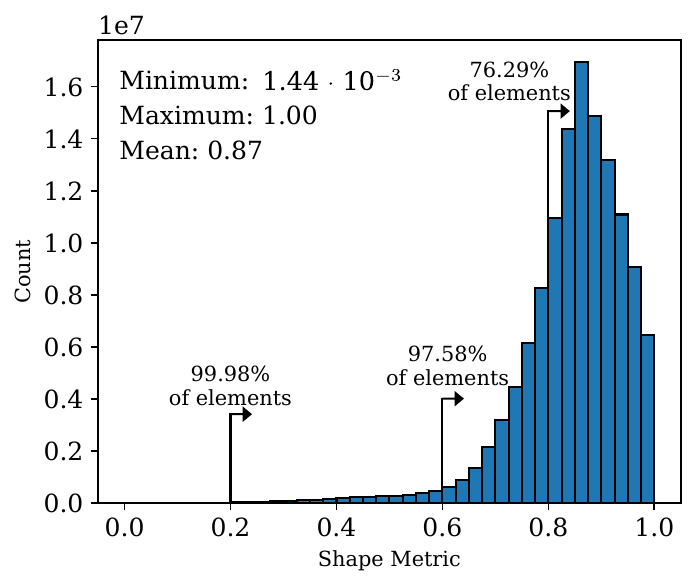}
        \caption{}
        \label{fig:ShapeFactor}
    \end{subfigure} 
    \begin{subfigure}{0.49\linewidth}
        \includegraphics[width=\linewidth]{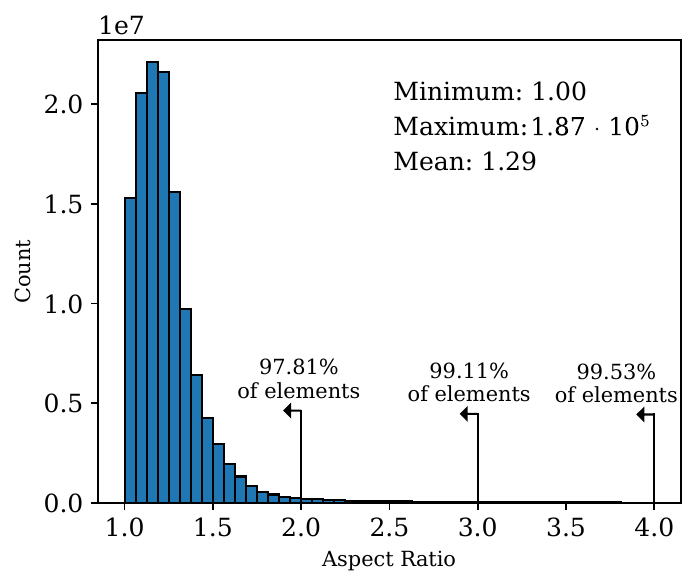}
        \caption{}
        \label{fig:ArHisto}
    \end{subfigure}
    
    \begin{subfigure}{0.49\linewidth}
        \centering
        \includegraphics[width=\linewidth]{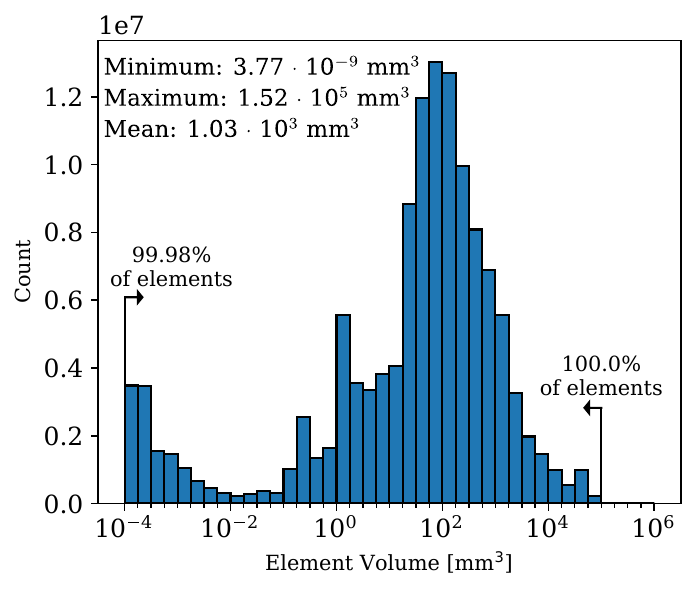  }
        \caption{}
        \label{fig:ElemHisto}
    \end{subfigure}
    \caption{Bar charts showing (a) the distribution of shape factor values for elements in the MAST-U mesh, (b) the distribution of aspect ratio values for elements in the MAST-U mesh, and (c) the distribution of element volumes for elements in the MAST-U mesh.}
    \label{fig:Histos}
\end{figure}

Element size is not the only factor that should be considered when selecting and evaluating an FEM mesh. The quality of the elements is also important. It is often desirable that elements have good shape properties, as poorly shaped elements can give rise to ill-conditioned matrices that tend to slow or even prevent convergence of iterative solvers \cite{KNUPP2003217}. There are numerous metrics used in literature to quantify mesh quality. Figure \ref{fig:Histos} shows the distributions of two mesh quality metrics for, and the volumes of, the elements in the final MAST-U mesh. The mesh quality metrics used are the shape factor derived by Knupp {\it et al.} \cite{KNUPP2003217} and the aspect ratio derived by Parthasarathy {\it et al.} \cite{PARTHASARATHY1994255}. The aspect ratio is defined as
\begin{equation}
    \text{Aspect Ratio} = \frac{\text{Circumsphere radius}}{3*\text{Inradius}},
    \label{eq:aspect_ratio}
\end{equation}
where the circumsphere radius and inradius are calculated for the tetrahedron being evaluated. The shape factor is defined as
\begin{equation}
    \text{Shape factor} = \frac{3}{\text{Mean ratio of weighted jacobian matrix}}
    \label{eq:shape_factor}
\end{equation}
using the mean ratio of the weighted jacobian matrix defined by Knupp \emph{et al.} \cite{KNUPP2003217}.


 Cubit's satisfactory value range for the shape factor is between 0.2 and 1. A value below 0.2 indicates that an element is of poor quality. Figure \ref{fig:ShapeFactor} shows that while there are elements within the mesh below this threshold, and thus of poor quality, the poor quality elements are vastly outnumbered by elements of acceptable quality, with the majority of elements having shape factor values of more than 0.8. A similar rationale can be applied to the aspect ratio values seen in Figure \ref{fig:ArHisto}. Cubit's acceptable range for the aspect ratio is between 1 and 3. There exist in the final mesh elements with aspect ratios above 3. In particular, there were 10 elements having a particularly poor aspect ratio, above 8000, that constitute less than 0.00001\% of the elements in the mesh. Despite the existence of elements with aspect ratios higher than 3, they are uncommon compared to elements of satisfactory aspect ratio. In a mesh with as many constituent parts as the MAST-U mesh, it is difficult to entirely eliminate poor quality elements. Meshing thousands of individual pieces of geometry will almost inevitably result in the presence of some poor quality elements. Even if a user can afford to produce an exceedingly high fidelity mesh that completely eliminates all the poor quality elements, the impact of removing such a small section of elements on the final results is unclear, and is potentially unfavourable given the increase in computational resource that would be required to use such a mesh. What our convergence analysis shows is that the overwhelming majority of the $\approx$ 127 million elements in the MAST-U FEM model are of acceptable quality and hence the final mesh is suitable for the analysis of mechanical behaviour of the full device, which we explore below.   

\subsection{Model Description and Future Improvements}
\subsubsection{Model Description} \label{ModelDescription}
The MAST-U finite element mesh used in this analysis involves 3,938 individual volumes, or parts, that are discretised into approximately 127 million tetrahedral elements. In this section we describe the steps involved in the development of the model, to show how the original model, which had previously been used for neutron transport calculations, has been upgraded to construct a model suitable for mechanical analysis. The neutron transport model is unsuitable for mechanical simulations since it lacks various mechanical supports and hence involves many floating components not compatible with conditions of mechanical equilibrium. There were also components absent from the neutron transport model, implying that the total mass of materials in the model using for neutron transport calculations was significantly lower than the mass of the real MAST-U device. To address this, the missing components were identified, modelled and implemented in the CAD, and subsequently in the FEM model.
\begin{figure}[t]
    \centering
    \begin{subfigure}[t]{0.49\linewidth}
        \includegraphics[width=0.9\linewidth]{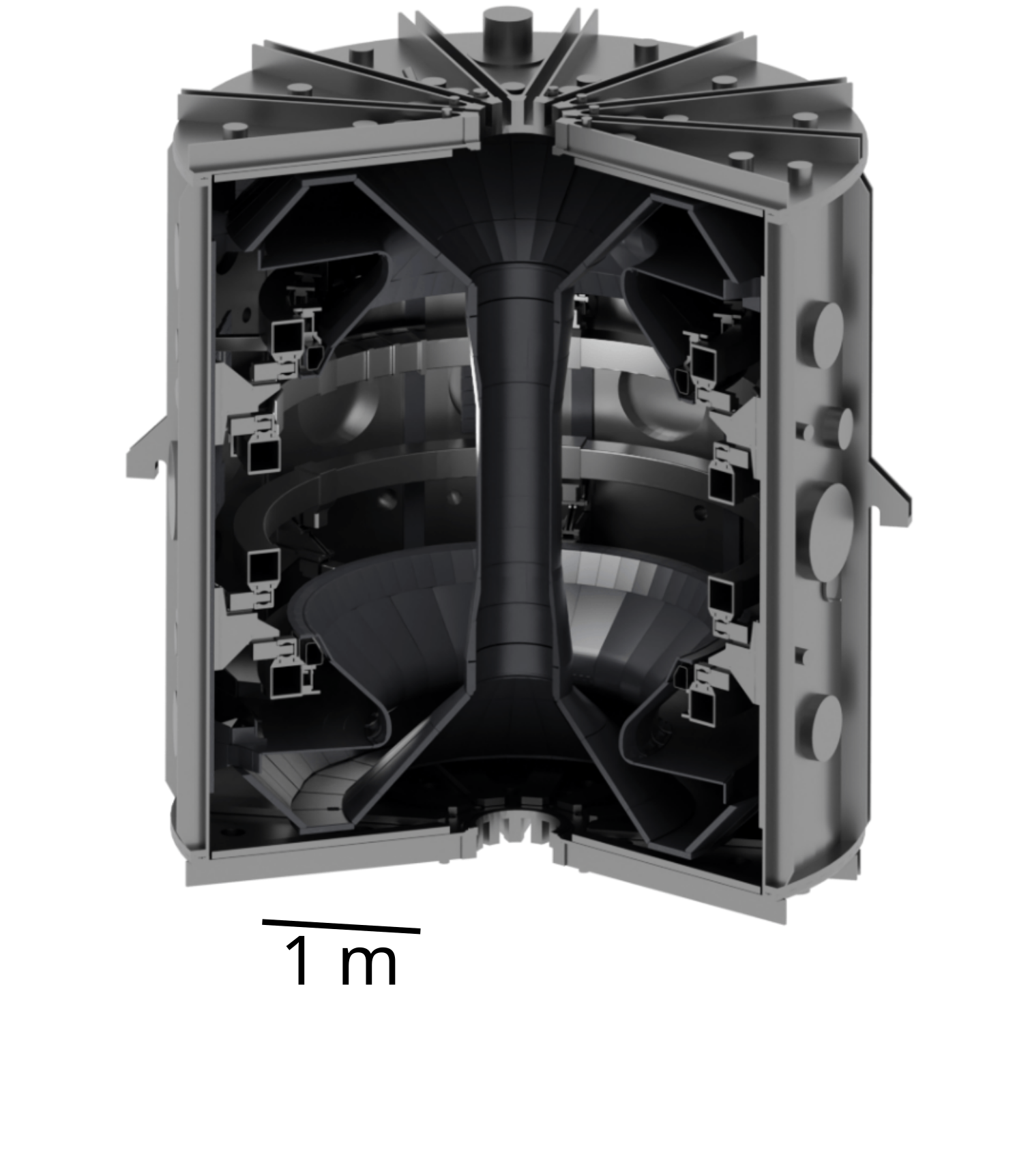}
        \caption{}
        \label{OriginalCAD}
    \end{subfigure}
    \begin{subfigure}[t]{0.49\linewidth}
        \includegraphics[width=0.9\linewidth]{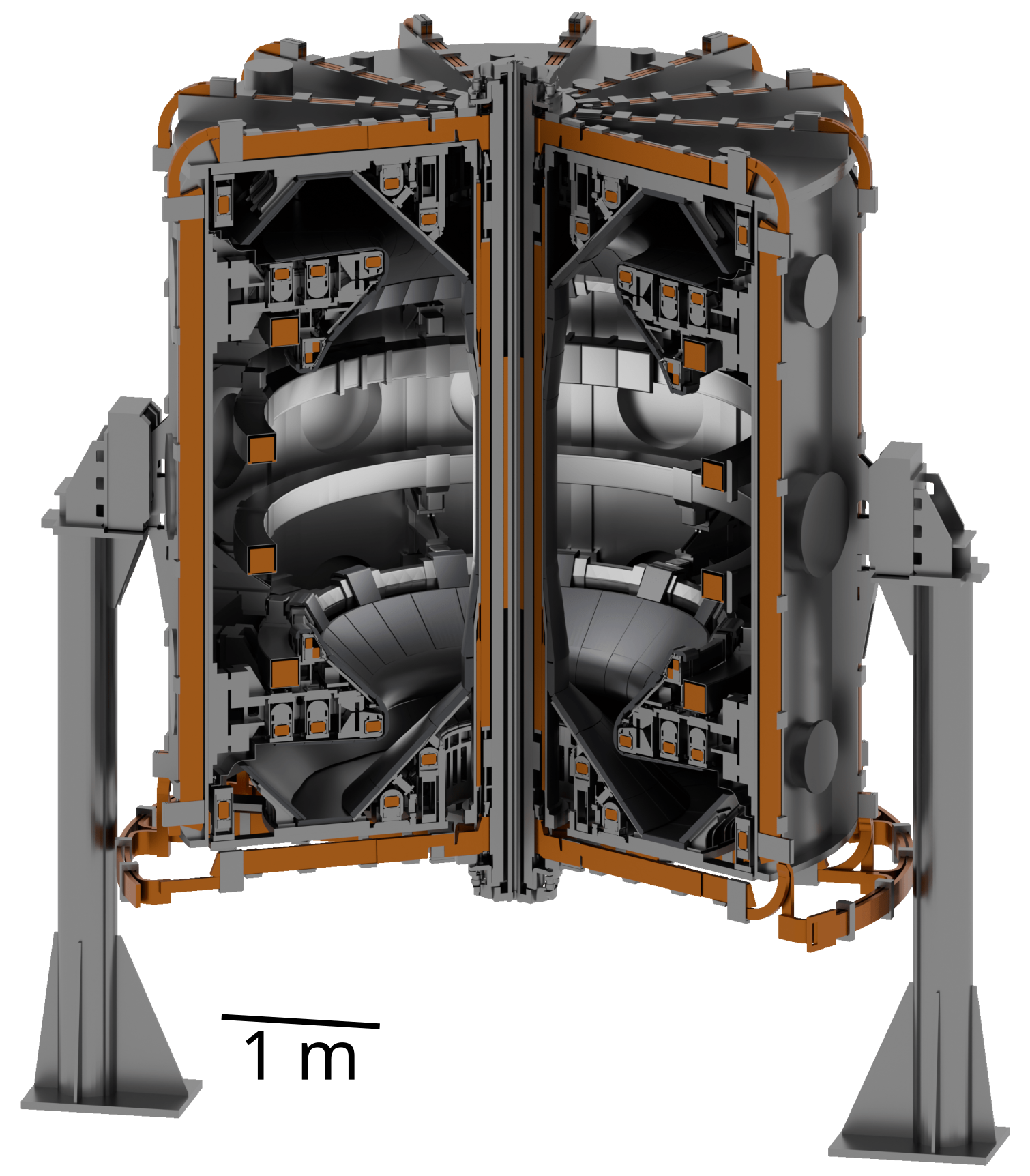}
        \caption{}
        \label{NewCAD}
    \end{subfigure}
    \caption{Comparison between (a) the initial model developed for neutron transport calculations, and (b) the final MAST-U CAD model providing input to the FEM analysis.}
    \label{fig:neutronics_model}
\end{figure}

\begin{figure}[t]
    \includegraphics[width=0.88\linewidth]{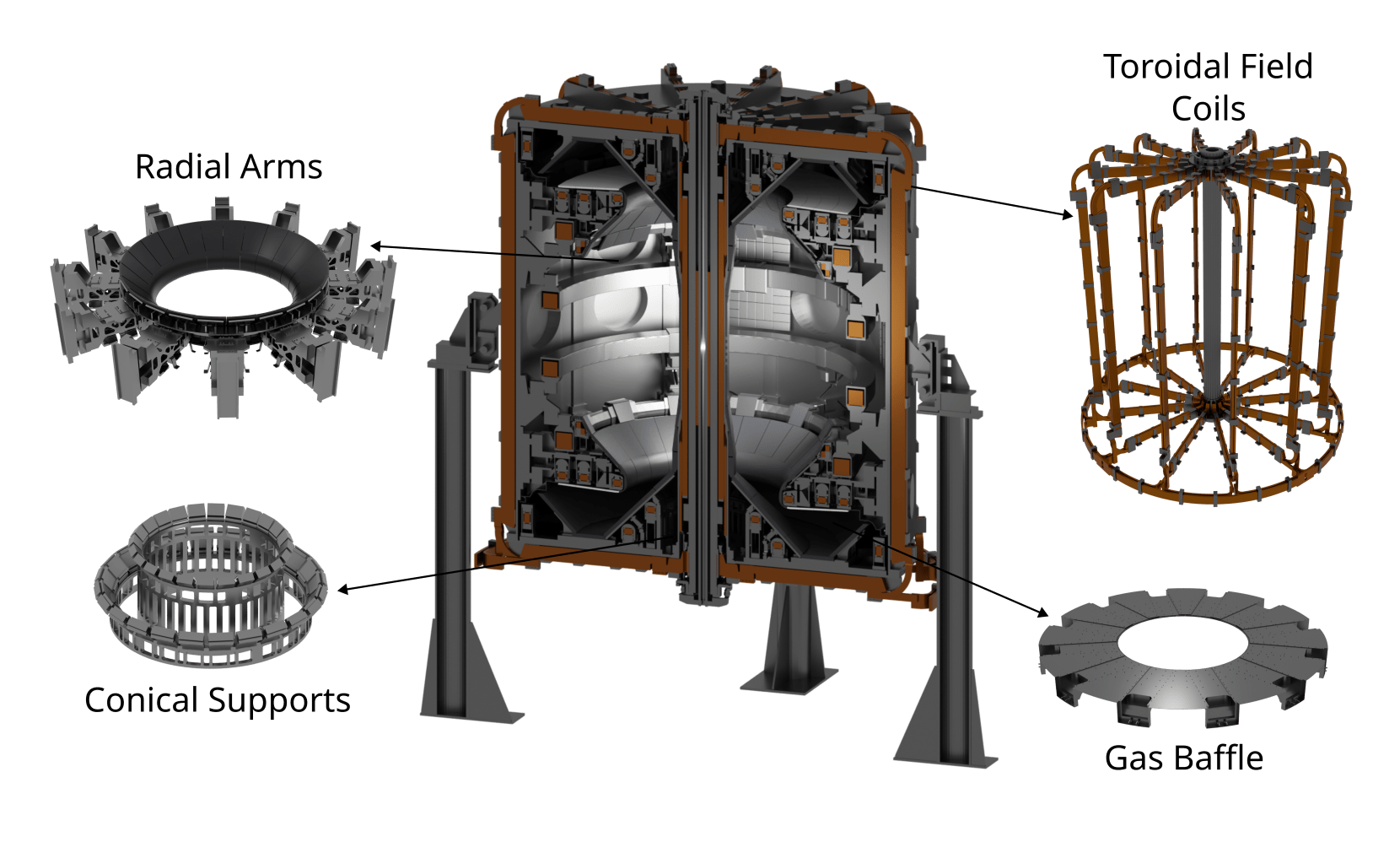}
    \caption{Full CAD assembly with annotation showing the placement of the radial arms, conical supports, gas baffle and toroidal field coils. Insets are not to scale.}
    \label{fig:MastUParts}
\end{figure}

Figure \ref{fig:neutronics_model} compares a sectioned view of the original model used for neutron transport analysis with a sectioned view of the final model. The original model already included the vacuum vessel, the vessel endplates, the casings for the poloidal field (PF) coil magnets, the inboard shielding for the vessel and the divertor panels. The first new components included in the model developed here were those necessary for supporting the MAST-U divertor nose tiles. Each nose tile is attached to a mount. All of the mounts are  attached to two metal rings, which themselves are supported by radial arms. Figure \ref{fig:MastUParts} shows the CAD models of these parts. The radial arms not only provide support to the gas baffle rings, but also support the P6 magnet support brackets \footnote{Here and below, the notations refer to the UKAEA Technical Document Management System (TDMS)}. Each radial arm is mounted to the inner wall of the vacuum vessel, which is supported by four legs attached to the cylindrical vacuum vessel by means of flexible load-bearing brackets. The gas baffle provides support for the legs of the radial arms as well as a mounting point for the gas baffle panels, which constitute a large section of the first wall shielding. Another section of the first wall requiring mechanical support were the conical shielding panels. In the real MAST-U tokamak, these panels are supported by steel panels which are mounted to two internal support rings. The modelled version of these components can be seen in Figure \ref{fig:MastUParts}.

\begin{figure}[h!]
    \begin{minipage}{0.48\linewidth}
    \begin{subfigure}[t]{\linewidth}
        \includegraphics[width=0.35\linewidth]{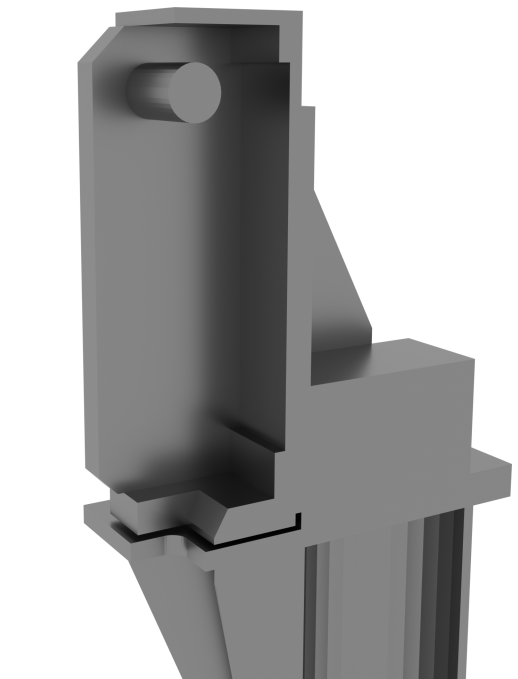}
        \includegraphics[width=0.2\linewidth]{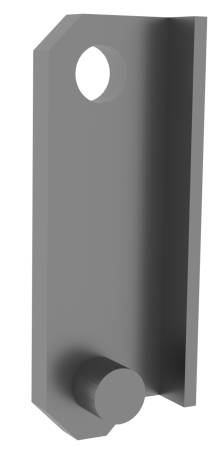}
        
        \includegraphics[width=0.4\linewidth]{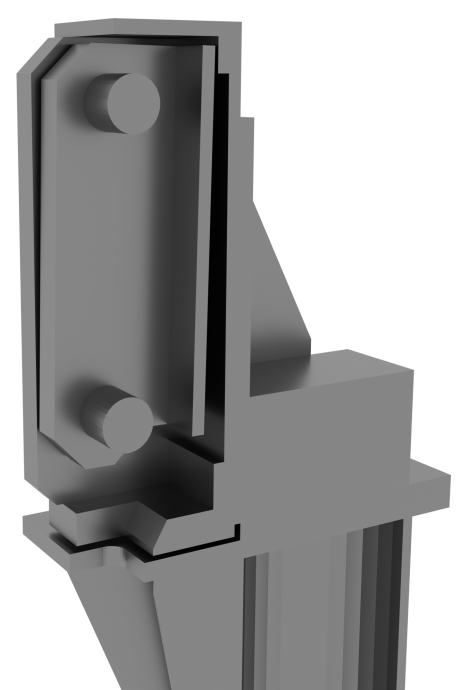}
        \caption{}
        \label{fig:MastULegSwing}
    \end{subfigure}
    \end{minipage}
    \begin{minipage}{0.48\linewidth}
    \begin{subfigure}[t]{\linewidth}
        \includegraphics[width=0.5\linewidth]{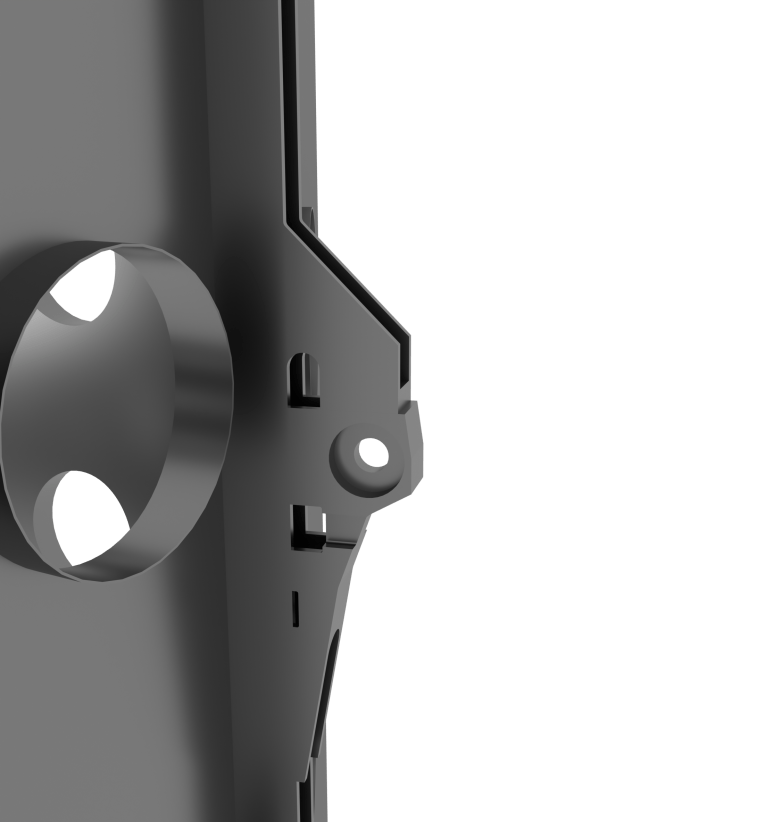}
        
        \includegraphics[width=0.5\linewidth]{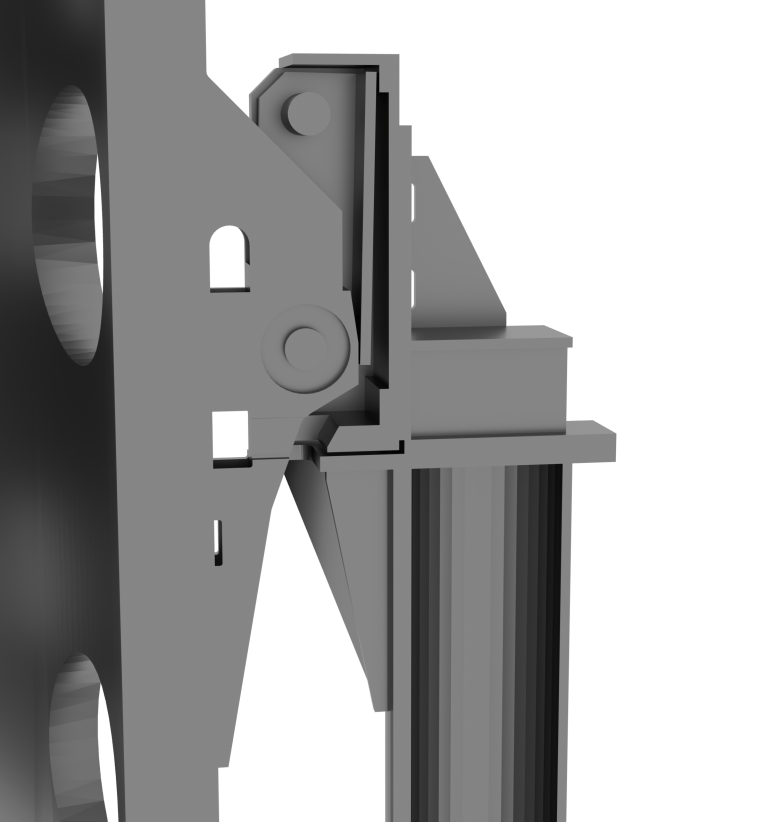}
        \caption{}
        \label{fig:MastULegBracket}
    \end{subfigure}
    \end{minipage}
    \caption{Figures illustrating the supporting leg attachment mechanism. (a) A clipped view of the components making up the MAST-U leg swing, showcasing the outer (top left), inner (top right), and combined assembly (bottom). (b) The vacuum vessel leg attachment bracket attached to the outside of the vaccum vessel (top) and a clipped view of the final leg attachment assembly with the vacuum vessel bracket mounted on the leg swing (bottom).}    
    \label{fig:LegAttachment}
\end{figure}

An important addition to the present mechanical model for the device, absent in the original neutron transport model but essential in the context of mechanical analysis, are the supporting legs. Given that these are the sole means of support for the vacuum vessel and its internal components, the stress values in the close proximity to the leg attachments are likely to be high. Therefore, the mechanism of attachment of the legs to the vacuum vessel needs to faithfully mimic the real tokamak device in order to ensure that the computed stress values are representative of reality. Figure \ref{fig:LegAttachment} details the leg attachment mechanism.
The method of attachment of the vacuum vessel to the legs is designed to accommodate vibrations of the device during operation and has similarities to a pendulum. The leg swings shown in Figure \ref{fig:MastULegSwing} are composed of an inner and outer parts. The inner part has a circular extrude that mounts to the pin of the outer part. The inner also has a pin which threads through the circular extrude in the attachment bracket shown in Figure \ref{fig:MastULegBracket}. Through gravity, the vessel weight is transmitted to the upper pin of the leg swing. 

There are also many new components implemented in to the model that do not provide mechanical support to other components but are essential for the plasma operation of the tokamak. One such set of components are the toroidal field magnets that surround the vessel, as seen in Figure \ref{fig:MastUParts}. The amount of copper in these coils contributes significantly towards the total mass of the device, and the copper magnets are therefore necessary to implement and include in the mechanical analysis. Many of the poloidal field magnets absent from the initial model have also been added. These include the D2, D3, D5, D6, D7, DP, P1, PX and PC coils, as well as their respective installation and supporting brackets.

\begin{figure}[t]
  \includegraphics[width=0.99\columnwidth]{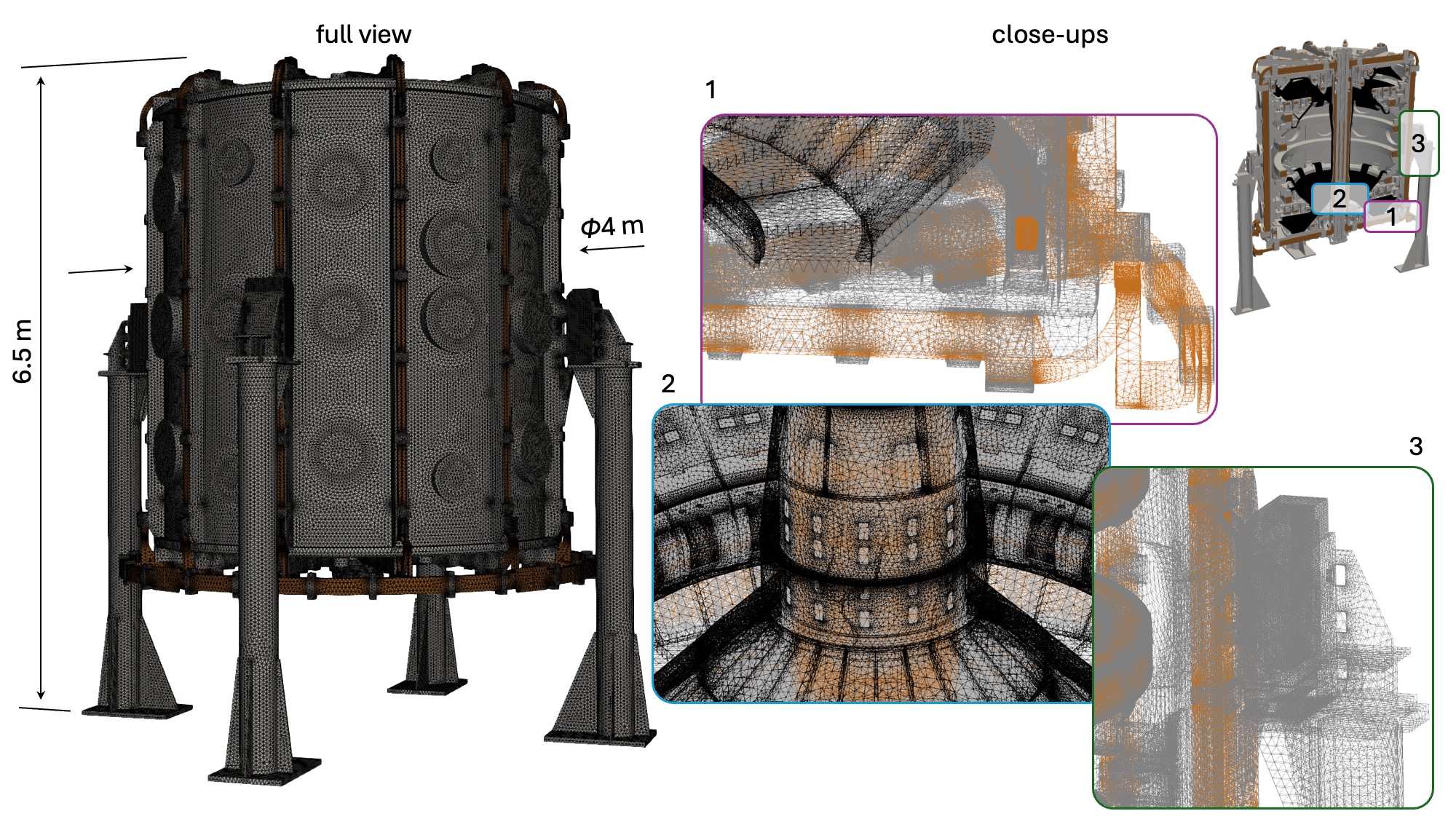} %
    \caption{Schematic view of a finite element method (FEM) representation of the  MAST-U spherical tokamak. The full FEM model involves in excess of 127 million individual tetrahedron elements, or 83 million DOFs. The total height of the device, including the legs supporting the cylinder, is about 6.5 metres, where the cylindrical vacuum vessel has the radius of 2.0 m and the height of 4.5 m. The volume of the plasma is approximately 8 m$^3$.  }
    \label{fig:MASTU_sketch}
\end{figure}

\begin{table}
\centering
\caption{Material property values used for materials in the MAST-U model.}
\label{Table:MaterialProperties}
\begin{tabular}{c|c|c|c} 
Material & Mass density [kg/m$^3$] & Young's modulus [GPa] & Poisson's ratio \\
\hline
Stainless steel & 7800 & 210 & 0.3 \\ 
Copper & 8960 & 125 & 0.3 \\ 
Graphite & 2260 & 11 & 0.3 \\ 
\end{tabular}
\end{table}
The parts in the MAST-U model are represented by three different structural materials: stainless steel, copper, and graphite. The material properties used for these materials in the analysis are shown in Table \ref{Table:MaterialProperties}. We note that the chosen values are more typical of a ferritic steel even if the vacuum vessel is made of stainless steel; the difference is however only of about 5\%. Using these material properties, we can compare the mass of the model with and without the additional components implemented for this analysis. Table \ref{table:MastMass} lists values of volume and mass for the different materials present within the MAST-U model. Most notable here is the increase in the total mass resulting from the new components modelled for this analysis. An overview of the final mesh is presented in Figure \ref{fig:MASTU_sketch}.
\begin{table}
\centering
\caption{Volume and mass values for the components in the starting MAST-U neutronics model, and the new model used for mechanical analysis. Mass and volume values are grouped by their respective material. Note that there were no copper components in the neutronics model, hence some entries in the table are not applicable.}
\label{table:MastMass}
\begin{tabular}{>{\centering\arraybackslash}m{3cm}|>{\centering\arraybackslash}m{3cm}|>{\centering\arraybackslash}m{3cm}|>{\centering\arraybackslash}m{3cm}|>{\centering\arraybackslash}m{3cm}} 
Material & Neutronics model volume [m$^3$] & Mechanical model volume [m$^3$] & Neutronics model mass [kg] & Mechanical model mass [kg]\\
\hline
Stainless steel & 3.73 & 8.45 & 29148 & 65936 \\ 
Copper & n/a & 2.93 & n/a & 26242 \\ 
Graphite & 1.48 & 1.75 & 3354 & 3946 \\ 
\hline
Total & 5.21 & 13.13 & 32502 & 96124 \\ 
\end{tabular}
\end{table}
\subsubsection{Limitations and Future Developments}
When generating a finite element model, there are still approximations that must be made to help avoid unwieldy element counts. Imprinting and merging surfaces is one such practice. It allows components to be bonded together by merging surfaces with like topology. This ensures the conformity of finite elements across contiguous component boundaries. In a perfect model, every bolt and screw would be modelled with high fidelity and contact mechanics would be used to model the physical interactions between every screw and threaded insert. While this might not be seen as a limitation, running a simulation for a model of such high fidelity would be infeasible given the current computational capabilities. The preparation time for constructing such a model would also be immense, whereas an excessive level of fidelity might also be unnecessary when investigating the larger scale characteristics of an entire tokamak device.

The limitation that became most apparent during the model's development was a human being having to be doing a lot of the preparation. Imprinting and merging surfaces and generating a mesh should theoretically be automatable. However, tolerance issues within CAD assemblies currently prevent this from being an established practice. Tolerance issues could result from components crossing CAD kernel boundaries, small overlaps resulting from poor alignment of components within CAD assemblies, or maybe large scale systemic volume overlaps in the original CAD due to its primary use case being something other than analysis, for example visualisation. Regardless of the cause, the presence of tolerance issues results in the need for significant human intervention when generating large scale models. How current CAD practices can be improved on this front is unclear. Perhaps parametric generation of large CAD models could help avoid geometry overlaps. What is clear is that the computational capability to perform analyses on large scale FEM models is presently available, and if the main impediment to the usage of such models continues to be the human time required to prepare them, then this point requires attention. Future studies might look to utilise the upcoming iso-geometric analysis (IGA) tools. IGA utilises the spline functions used for defining CAD geometry as basis functions for finite element analysis \cite{IGA}. This methodology enables accurate representations of the CAD geometry in finite element solvers, as well as a rich function space for representing the degrees of freedom.

\section{Structural calculations}
Elastic stress will build up in a fusion reactor due to several different reasons. Even if the machine is not operational, gravitational forces are always active. During operation, magnetic forces will be present where current is passed through the materials and also in the ferromagnetic materials such as ferritic steels. Inevitably, transient and steady temperature variations will produce thermal loads due to differential thermal expansion. Finally, in the case of power plants, prolonged exposure to irradiation will induce location-dependent property and dimensional variations such as swelling. In this study, we consider only some of these loads. For instance, neither ferromagnetic materials nor intense heat fluxes nor irradiation effects are present in the MAST-U tokamak. The objective of the structural analysis performed below is therefore to compute the displacement field $u_i(\mathbf{x})$ given all the applied loads, subject to boundary conditions. For reference, the static calculations presented below utilised about 500 CPU hours each. Running on about 600 to 800 CPUs in parallel, they took less than one hour to complete.

The equation of motion governing the components of the time-dependent displacement field ${\bf u}(\mathbf{x},t)$ in a linear elastic body is \cite{LandauElasticity}
\begin{equation}\label{eq:motion}
    c_{ijkl}u_{k, lj}(\mathbf{x},t)+f_i(\mathbf{x},t)=\rho\Ddot{u_i},
\end{equation}
where $c_{ijkl}$ is the fourth-rank tensor of elastic stiffness constants, $f_i({\bf x},t)$ is the body force acting on the material at point ${\bf x}$, and $\rho$ is the mass density. Suffix notation is used with the convention that indexes after a comma imply differentiation with respect to spatial coordinate with the corresponding index. In general, the materials can experience thermal expansion and irradiation-induced swelling, and could be subject to forces of gravitational and electromagnetic nature. All this can be captured by suitable distributions of body forces complemented by tractions at the boundary of the structure, see Ref. \cite{Reali2024} for further detail.

In mechanical equilibrium, the right-hand side of Eq.~\eqref{eq:motion} vanishes and components of the displacement field obey the linear differential equation \cite{Reali2024} 
\begin{align}\label{eq:PDE_aniso}
    c_{ijkl}\Big[u_{k, lj}(\mathbf{x})-\alpha_{kl}\frac{\partial T(\mathbf{x})}{\partial x_j}&-\Omega_{kl, j}(\mathbf{x})\Big] 
    + \rho(\mathbf{x})g_i+f_i^m(\mathbf{x})=0.
\end{align}
Here, $\alpha_{kl}$ is the thermal expansion coefficients tensor, $T$ is the temperature relative to the thermally undeformed material, $\Omega_{kl}$ is the dimensionless swelling tensor, or equivalently the irradiation-induced eigenstrain, $\rho$ is the density, $g_i$ the gravitational acceleration and $f_i^m$ the magnetic body force. Swelling stems from the accumulation of defects in the material under irradiation, and function $\Omega_{kl}({\bf x})$ equals the density of relaxation volumes of radiation defects at ${\bf x}$ \cite{Dudarev2018,Reali2022}.  
Under the simplifying assumptions of isotropic elasticity for $c_{ijkl}$ and isotropic thermal expansion and swelling, Eq.~\eqref{eq:PDE_aniso} can be reduced to
\begin{align}\label{eq:PDE_iso}
    \frac{\mu}{1-2\nu}u_{k,ik} + \mu u_{i, kk}-&2\mu\alpha\frac{(1+\nu)}{(1-2\nu)}\frac{\partial T}{\partial x_i} 
    -\frac{2\mu}{3}\frac{(1+\nu)}{(1-2\nu)}\frac{\partial \Omega}{\partial x_i}  +\rho g_i + f_i^m = \rho\frac{\partial^2 u_i}{\partial{t^2}},
\end{align}
where $\mu$ and $\nu$ are shear modulus and Poisson's ratio. Eqs.~\eqref{eq:PDE_aniso} and \eqref{eq:PDE_iso} are stated here for completeness, as they are foundational for further developments \textemdash additional details about the elastic analysis and the swelling tensor $\Omega_{kl}$ can be found in Refs.~\cite{Reali2024,Dudarev2018} \textemdash but thermo-magnetic loads as well as radiation swelling are not treated in this study.

The elastic equilibrium condition can be solved for geometrically complex structures using FEM. Here, the tensor mechanics module of the highly parallel code MOOSE was used \cite{Gaston2009, Permann2020}. If Eq.~\eqref{eq:PDE_aniso} is a linear equation, the numerical framework can include various sources of nonlinearity. The commonly considered ones are the contact between adjacent bodies, large strains beyond the small strain limit, and nonlinear behaviour of materials such as plastic deformation and thermal or irradiation creep. 

If, on the other hand, we neglect body forces and effects of thermal expansion and swelling but include the inertial dynamic term of Eq.~\eqref{eq:motion} in the limit of isotropic elasticity we arrive at \cite{LandauElasticity}
\begin{equation}\label{eq:dynamic}
    \frac{\mu}{1-2\nu}u_{k,ik} + \mu u_{i, kk}=\rho\frac{\partial^2 u_i}{\partial{t^2}},
\end{equation}

Investigating the dynamic response of the full tokamak device is especially important if seismic events are likely to occur or if magnetic forces are cyclic in nature, e.g. during high-frequency pulsed operations. Although the seismic analysis is not immediately relevant for MAST-U tokamak in the UK, the matter is doubtlessly significant for ITER in France \cite{Schioler2011, Bachmann2017}, and possibly for some of the future locations of DEMO. As described in the documentation for the MOOSE software \cite{Permann2020}, taking a one-dimensional case for simplicity, a frequency domain calculation can be performed as follows. An equation of motion 
\begin{equation}\label{eq:1Ddynamic}
    \mu \frac{\textnormal{d}u}{\textnormal{d}x^2}=\rho\frac{\textnormal{d}u}{\textnormal{d}t^2}
\end{equation}
that follows from (\ref{eq:dynamic}) if we assume that displacements are orthogonal to the direction of variation of $u(x,t)$, can be cast in the frequency domain by making the ansatz of a purely oscillating solution $u(x,t)={\tilde u}(x)\exp{(i\omega t)}$. Taking the time Fourier transform of $u(x, t)$, we arrive at an equation for transverse waves in an elastic material
\begin{equation}\label{eq:1DFourier}
    \mu \frac{\textnormal{d}\tilde{u}}{\textnormal{d}x^2}+\rho\omega^2\tilde{u}=0.
\end{equation}
Similarly, a general form of equation (\ref{eq:dynamic}) in the frequency domain is
\begin{equation}\label{eq:dynamic_frequency}
    \frac{\mu}{1-2\nu}\tilde u_{k,ik} + \mu \tilde u_{i, kk}+\rho\omega ^2 {\tilde u_i}+\tilde f_i=0,
\end{equation}
where $\tilde f_i({\bf x},\omega)$ is a time Fourier transform of the body force. Solutions of homogeneous equation \eqref{eq:dynamic_frequency} of the form 
$\tilde u_i(x)=\tilde u^0_i\exp(i{\bf k}\cdot {\bf x})$ describe sound waves with dispersion relations $\omega (k)=k\sqrt{\mu/\rho}$ for the shear waves where the displacements are orthogonal to the direction of propagation of the wave, and $\omega(k)=k\sqrt{2\mu (1-\nu)/\rho (1-2\nu)}$ for the longitudinal compression waves similar to sound waves propagating in a liquid. The fact that equation (\ref{eq:dynamic_frequency}) is linear and similar to equation (\ref{eq:PDE_iso}) describing mechanical equilibrium, enables applying large-scale FEM simulations to the investigation of the response of complex structures to dynamic perturbations \cite{Bunting2020}. We note here that $\omega$ denotes the so-called angular frequency that must not be confused with conventional frequency ${\sf f}$, expressed in cycles per second and measured in Hertz. The latter is related to the former by $\omega=2\pi {\sf f}$. In Section \ref{sec:freq} we simulate the dynamic response of MAST-U in the frequency domain, using the same FEM model that we use for computing the stress and strain fields in the tokamak structure under static gravitational and atmospheric pressure loading. 

\subsection{The density of stress}
As the complexity of FEM models increases, it is desirable to find a way to represent the state of stress predicted by a large-scale simulation in a simple way and that enables comparing either stresses of different physical origin, or their time evolution. This is why we adapt a quantity, known in the theory of electronic structure as the density of states, and introduce here the notion of the \emph{density of stress} (DOS), $D(\sigma)$. Given a body or collection of bodies of total volume $V$, and the stress as a function of position $\sigma(\mathbf{x})$, the DOS represents the volume fraction that experiences stress in the interval between $\sigma$ and $\sigma+\delta\sigma$. Using the probability distribution of stress is technologically highly relevant as the failure probability can be defined in the most straightforward manner as the overlap integral between the stress and strength distributions. The DOS may be defined for any of the components of the stress tensor, or some other characteristic of stress, say the von Mises stress 
$$
\sigma_{vM}=\sqrt{{3\over 2} \left(\sigma_{ij}\sigma_{ji}-{1\over 3}\sigma_{ii}\right)}=\sqrt{{1\over 2}\left[(\sigma_{11}-\sigma_{22})^2 + (\sigma_{22}-\sigma_{33})^2 + (\sigma_{33}-\sigma_{11})^2 +6(\sigma_{23}^2+\sigma _{31}^2+\sigma_{12}^2)\right]},$$
or the hydrostatic stress $\sigma_h=\sigma_{ii}/3=(\sigma_{xx}+\sigma_{yy}+\sigma_{zz}$)/3. We define the DOS as
\begin{equation}\label{eq:DOS_definition}
    D(\sigma)=\frac{1}{V}\int_V \text{d}\mathbf{x}^3 \ \delta\left(\sigma-\sigma(\mathbf{x})\right),
\end{equation}
using the Dirac delta function $\delta({\bf x})$. This is a generalised function, defined as $\delta({\bf x})=0$ for ${\bf x}\ne 0$, and ${\int f(\mathbf{x})\delta(\mathbf{x}-\mathbf{y}) \text{d}^3\mathbf{x}=f(\mathbf{y})}$. In our case $\int_{-\infty}^\infty\delta(\sigma) \text{d}\sigma = 1$, which shows that the units of this function are, if stress is its argument, MPa$^{-1}$. We can interpret the DOS as the probability density of finding a given stress value anywhere in the volume of the structure, since the notion of probability refers to the volume fraction of the body that is under a chosen value of stress. 

We now provide two examples where this quantity can be calculated analytically. The first example is a pillar of height $h$ and density $\rho$ standing  on a flat surface, where gravity (acceleration $g$) is the only applied load. If the vertical direction is denoted by $z$ and the origin is at the bottom of the pillar, we have that $\sigma_{zz}=-\rho g (h-z)$, see \cite{LandauElasticity}. Since the stress varies linearly and only depends on $z$, intuitively we can see that every stress value has the same probability as any other to be observed, from the minimum value of $-\rho g h$ at the bottom to the value of zero at the top. Therefore we expect a flat DOS. By applying the definition we can calculate it explicitly:
\begin{equation}
    D(\sigma_{zz})=\frac{1}{h}\int_0^h\delta\left[\sigma_{zz}-\left(-\rho g (h-z)\right)\right]=\frac{1}{\rho g h}H(\sigma_{zz}+\rho g h)H(-\sigma_{zz}).
\end{equation}
This can be shown e.g. by introducing the change of variable $\sigma_{zz}+\rho g (h-z)=u$. The expression involves the Heaviside step-function $H(x)$, which evaluates to 1 if the argument is positive and to 0 if the argument is negative. This function arises when extending the range of integration to infinity to calculate the integral using the properties of the Dirac delta function. The two step-functions combined produce a uniform probability distribution spanning an interval of stress values from $-\rho gh$ to $0$.

\begin{figure}[t]
    \centering
\subfloat[]{%
  \includegraphics[width=.49\linewidth]{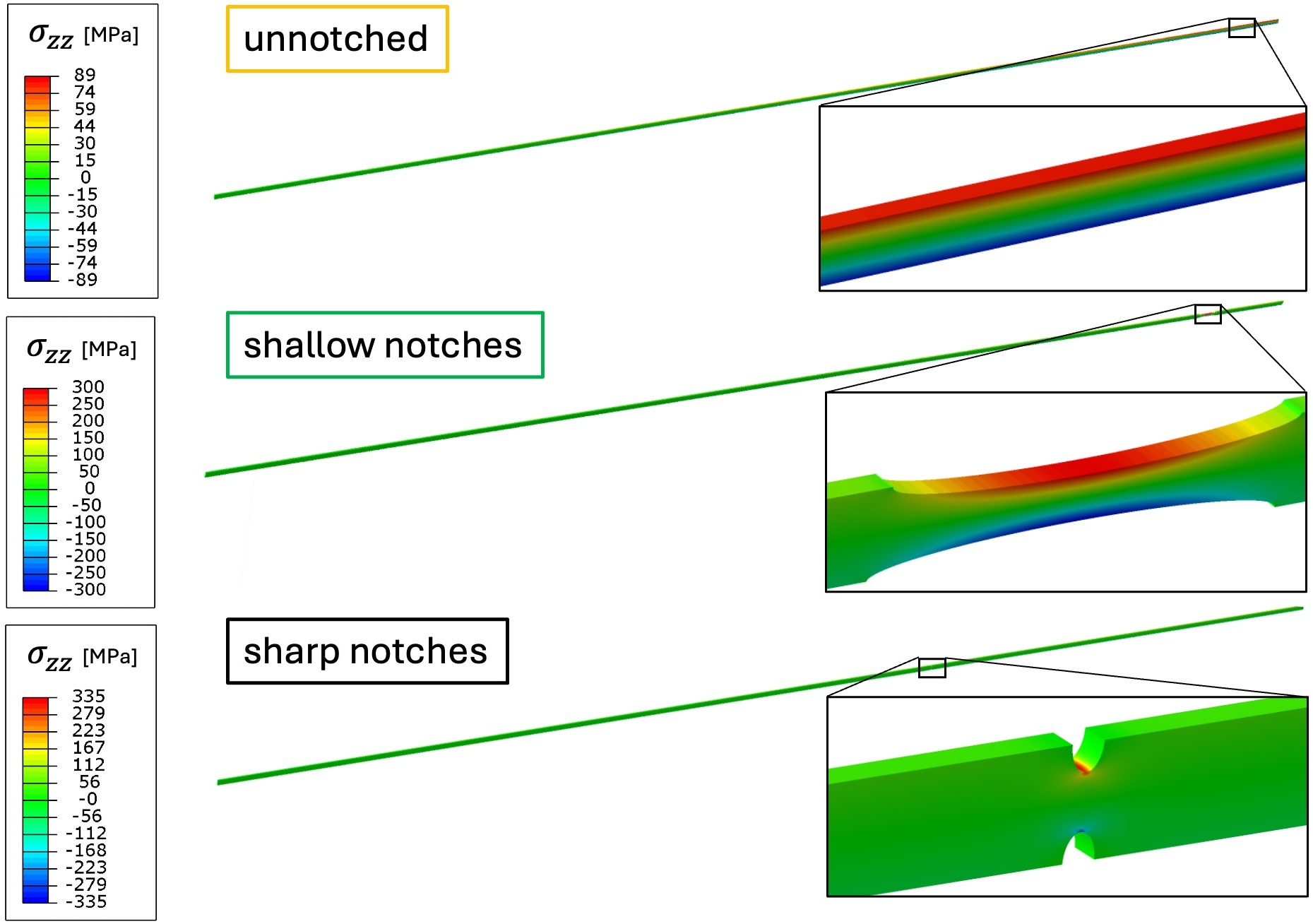}%
}\hfill    
\subfloat[]{%
  \includegraphics[width=.49\linewidth]{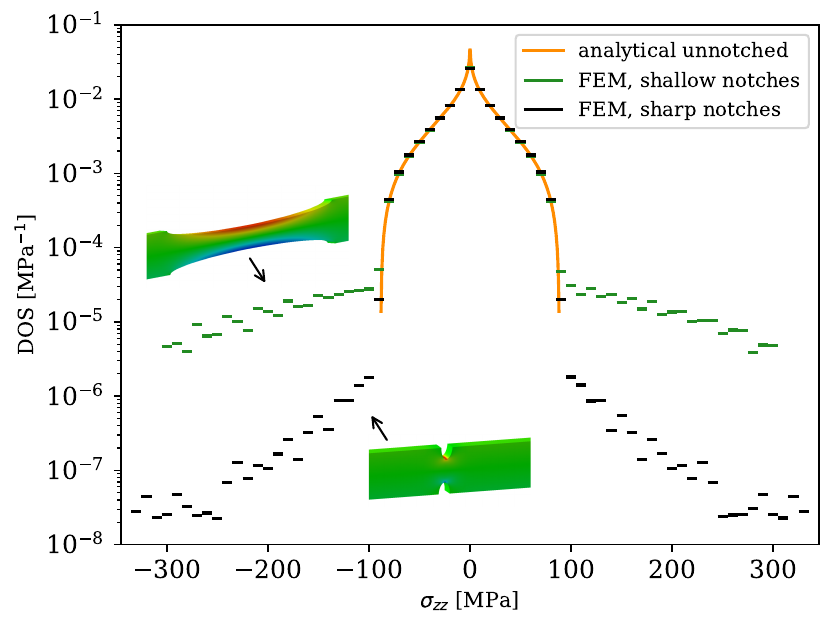}%
}\hfill    
\caption{(a) $\sigma_{zz}$ stress component in a cantilever beam aligned along the $z$ direction and loaded along $y$ by gravity and a concentrated force. The diagram at the top illustrates the case where there are no notches and the bending stress only depends on $y$. $\sigma_{zz}$ becomes localised if shallow (case in the middle) or sharp (case at the bottom) notches are introduced. (b) The density of stress for the beam with no notches is given by the analytical expression \eqref{eq:DOSbeam}. The DOS plots computed numerically show that a beam with a shallow or a sharp notch still exhibits the same stress distribution at large distances from the notches, but at the same time shows significant differences at extremes because of the notches that generate much higher stress in their vicinity. FEM computed stresses were binned over 10~MPa intervals, shown as lines of matching width.}
    \label{fig:beam}
\end{figure}
Another, less intuitive, example is explained in detail in App.~\ref{app:beam}. What is the probability distribution of stress in a cantilever beam? The beam is clamped at one end and is loaded by gravity and by a concentrated force at the free end corresponding to the entire weight of the beam. The beam has density $\rho$, length L along $z$ coordinate and a rectangular cross section of thickness $t$ in the vertical $y$ direction. The dominant stress component is the bending stress, which is maximum at the clamping position and has a known analytical expression. The maximum is $\sigma_{\text{max}}=9\rho g L^2/t$ and there is a symmetric tensile and compressive stress state. We show in App.~\ref{app:beam} that the DOS in this case is
\begin{equation}\label{eq:DOSbeam}
    D(\sigma_{zz})=\begin{cases}
			\frac{3}{4\sigma_{\text{max}}}\log\left[\frac{1+\sqrt{1-3\sigma/\sigma_{\text{max}}}}{3(\sqrt{1-3\sigma/\sigma_{\text{max}}}-1)}\right], & -\sigma_{\text{max}}\leq\sigma<0\\
			\frac{3}{4\sigma_{\text{max}}}\log\left[\frac{1+\sqrt{1+3\sigma/\sigma_{\text{max}}}}{3(\sqrt{1+3\sigma/\sigma_{\text{max}}}-1)}\right], & 0<\sigma\leq\sigma_{\text{max}}\\
            0, & \text{otherwise}
		 \end{cases}
\end{equation}
This distribution, as expected, extends between $+\sigma_{\text{max}}$ and $-\sigma_{\text{max}}$, it is symmetric in the tensile and compressive regions, and integrates to 1. The facts that it has (an integrable) singularity and that it decreases exponentially towards the highest stress values are the features that are not necessarily trivial or expected.

A large-scale FEM calculation involving $N$ elements produces a list of $N$ values of stress $\sigma_i$ \textemdash say the elemental stress at integration points \textemdash and a list of the corresponding volumes for each element, $V_i$. In practice, rather than computing a probability distribution using Eq.~\eqref{eq:DOS_definition}, we compute a normalised histogram. For every element, we sort its stress value into the correct bin. Second, we add the corresponding volume to the tally of that bin. This gives the total volume of all the elements whose stress can be sorted into any one bin. Figure \ref{fig:beam} gives a comparison between the analytically evaluated DOS given by Eq.~\eqref{eq:DOSbeam} and two numerical FEM examples. We considered a beam with length 500~mm, 2~mm thickness along the vertical direction and 0.5~mm thickness along the horizontal direction. The following material properties were assumed: $\rho=8000$~kg/m$^3$, $E=193$~GPa, $\nu=0.3$. We investigated how the distribution changes if either a shallow or a sharp symmetrical double notch were added to the clamped beam. The shallow notch was a semi-ellipse having a 5~mm major semi-axis along the beam direction and 0.5~mm minor semi-axis, centred 45~mm from the clamped end. The sharp notch was a semi-ellipse having a 0.25~mm minor semi-axis along the beam direction and a 0.5~mm major semi-axis, centred 200~mm from the clamped end. In both cases the notches removed less than 1\% of the material and only had a localised effect. Elsewhere the stress distribution in the beam was almost unaffected by the presence of the notches. Near the notches, however, the tensile and compressive stress could reach magnitudes well above $\sigma_{\text{max}}$. In the two simulations the notches were made in different locations, the sharp at a location with less bending moment, in order to have similar peak stresses. We note that the stress concentration around a sharp notch has a more localised effect than around a shallow notch. This is reflected in the slope of the DOS, which decays faster at extremes in the former than in the latter case.

\section{Full tokamak analysis}
The full reactor finite element model was tested using three different types of loads: gravitational, dynamic seismic and atmospheric pressure. Often, the size of FEM analyses can be reduced exploiting symmetries. This however requires the geometry, the loads and the boundary conditions to comply with the expected symmetry condition. The number of simulated volumes is halved for each of the symmetry planes that can be defined. In our case, the geometry is not perfectly symmetric because of the position of the apertures and other smaller details. Nonetheless, adopting some small approximations, the reactor structure exhibits a two-fold symmetry. Gravity and atmospheric pressure loads comply with this two-fold symmetry. The seismic analysis, however, does not. To have a flexible and general model that can be extended as the need arises, it is essential to retain the full geometry. This can only be enabled by a scalable and parallel HPC framework.

\subsection{Static deformation due to gravity}
Even in the absence of any load of thermal, mechanical or magnetic origin, and without irradiation, any large-scale device experiences significant gravitational pull, which in most cases is still less relevant than other loads. Therefore, the first test for our virtual tokamak model involved a simulation of its deformation under gravity. Only a comparison of a model with and without gravitational loads can ensure that the latter can indeed be disregarded. Moreover, gravity acts over the entire lifetime of the reactor and may become relevant in the presence of low-stress but persistent irradiation creep. This is not considered to be a concern for MAST-U due to the low levels of accumulated damage, but may become a significant issue for commercial fusion reactors \cite{Peacock2004,Luzginova2011}.

\begin{figure*}[t]
\subfloat[]{%
  \includegraphics[width=0.48\textwidth]{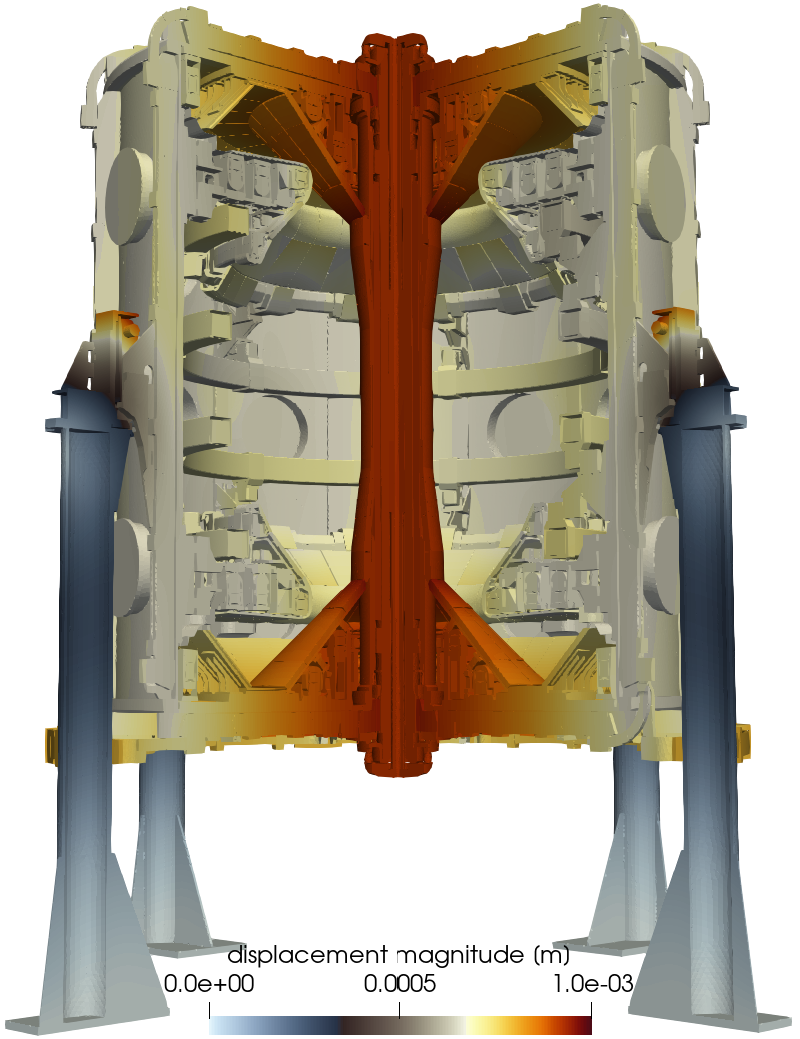}\label{fig:gravity_disp} %
}
\subfloat[]{%
  \includegraphics[width=.48\textwidth]{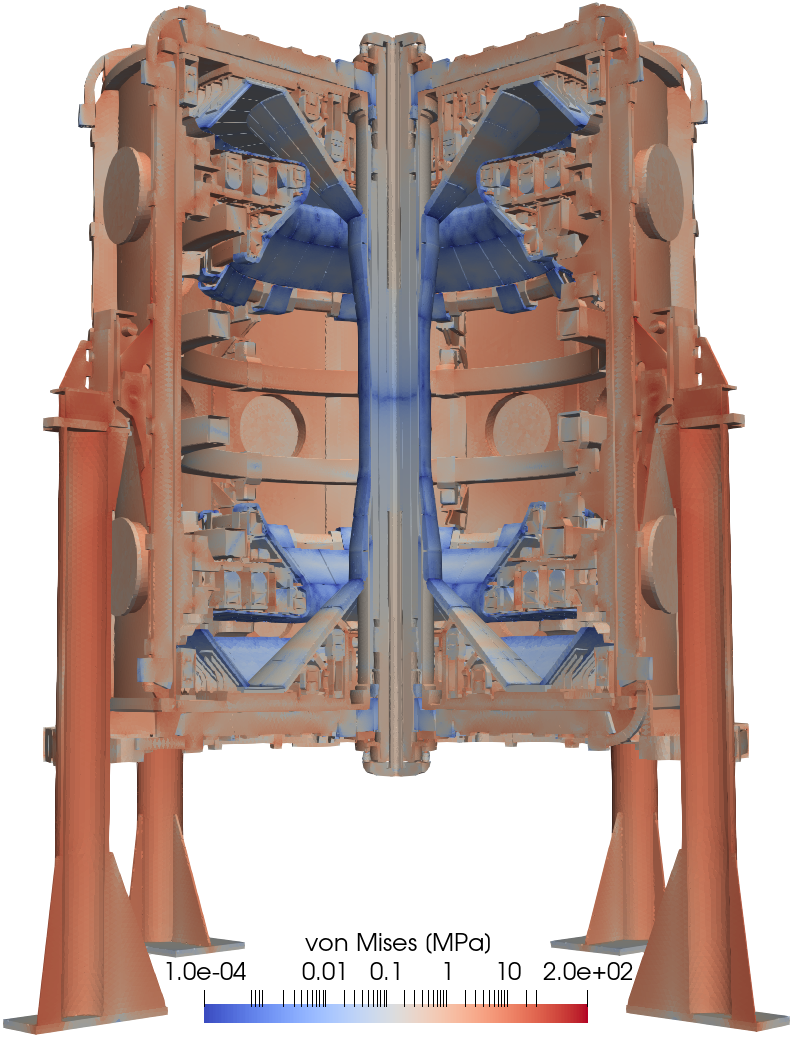}\label{fig:gravity_vM} %
}\hfill

\subfloat[]{%
  \includegraphics[width=.48\textwidth]{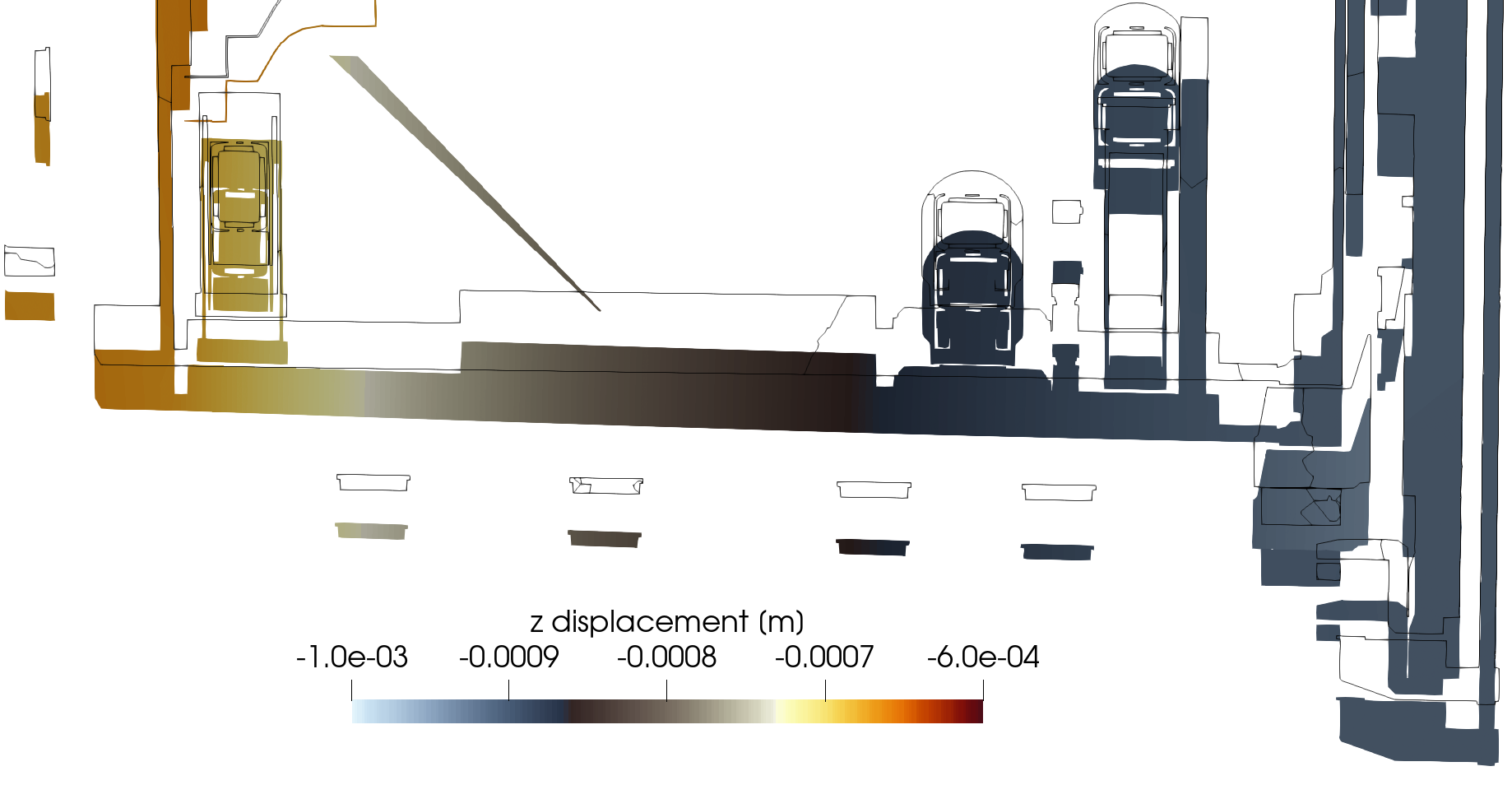}\label{fig:sagging} %
}\hfill
    \caption{MAST-U deformation and stress resulting from  gravitational forces. The weight of the device generates a downward displacement of the order of one millimetre, as shown in (a) by the displacement magnitude field and in (c) by the close-up view of the $z$ cross-sectional view of the bottom part of the vacuum vessel, where displacements were magnified by a factor of 100 to highlight the sagging induced by the weight of the central column. The resulting von Mises stress field is shown in (b) on a logarithmic scale, highlighting the fact that the load is highest at the four brackets on which the tokamak vessel is suspended.}
    \label{fig:gravity} 
\end{figure*}

For simulating the effect of gravity on the MAST-U structure, we applied a constant body force $\textbf{f}=\rho \textbf{g}$, where $\rho$ is the mass density, different for each material. Figure \ref{fig:gravity} shows the downward displacement of the structure and the resulting stress field in the full tokamak structure, taking the von Mises invariant as a representative quantity. The total mass of the vessel weighs on the four brackets, which slightly rotate, producing a horizontal displacement of about \SI{0.5}{\milli\meter}. The weight of the central column produces an additional sagging of about \SI{0.5}{\milli\meter} of the top and bottom faces of the vacuum vessel. Overall, we found a downward displacement of \SI{1}{\milli\meter}. Although highly accurate measures are not possible because of the structure that encloses the vessel, the available measured values of displacements in the device are close to a few millimetres.

The values of stress were overall modest, although at the supporting brackets the von Mises and the maximum principal stress reached values of about \SI{150}{\mega\pascal}. Excluding this region, the stresses were below \SI{50}{\mega\pascal}.

\subsection{Atmospheric pressure}\label{sec:pressure}
Fusion tokamaks operate under high vacuum. When they are not vented, the atmospheric pressure compresses the reactor from the outside with a pressure of \SI{101.3}{\kilo\pascal}. Similarly to the gravitational case, this is a load that is routinely disregarded in structural analyses of mechanical components. We run a second test where atmospheric pressure was applied to every external surface of the device. The results were in qualitative agreement with expectations: the vacuum vessel bends inwards on the sides and at the top/bottom surfaces, with the central column acting as a supporting pillar that prevents inwards deformation. Maximum displacements were of about \SI{0.2}{\milli\meter}, slightly lower than in the gravitational case probably due to the fact that the total force vanishes if pressure is integrated over all surfaces, which is not the case if gravitational body forces are integrated over the volume.

\begin{figure}[p]
\begin{minipage}{1.\textwidth}
  \includegraphics[width=0.95\textwidth]{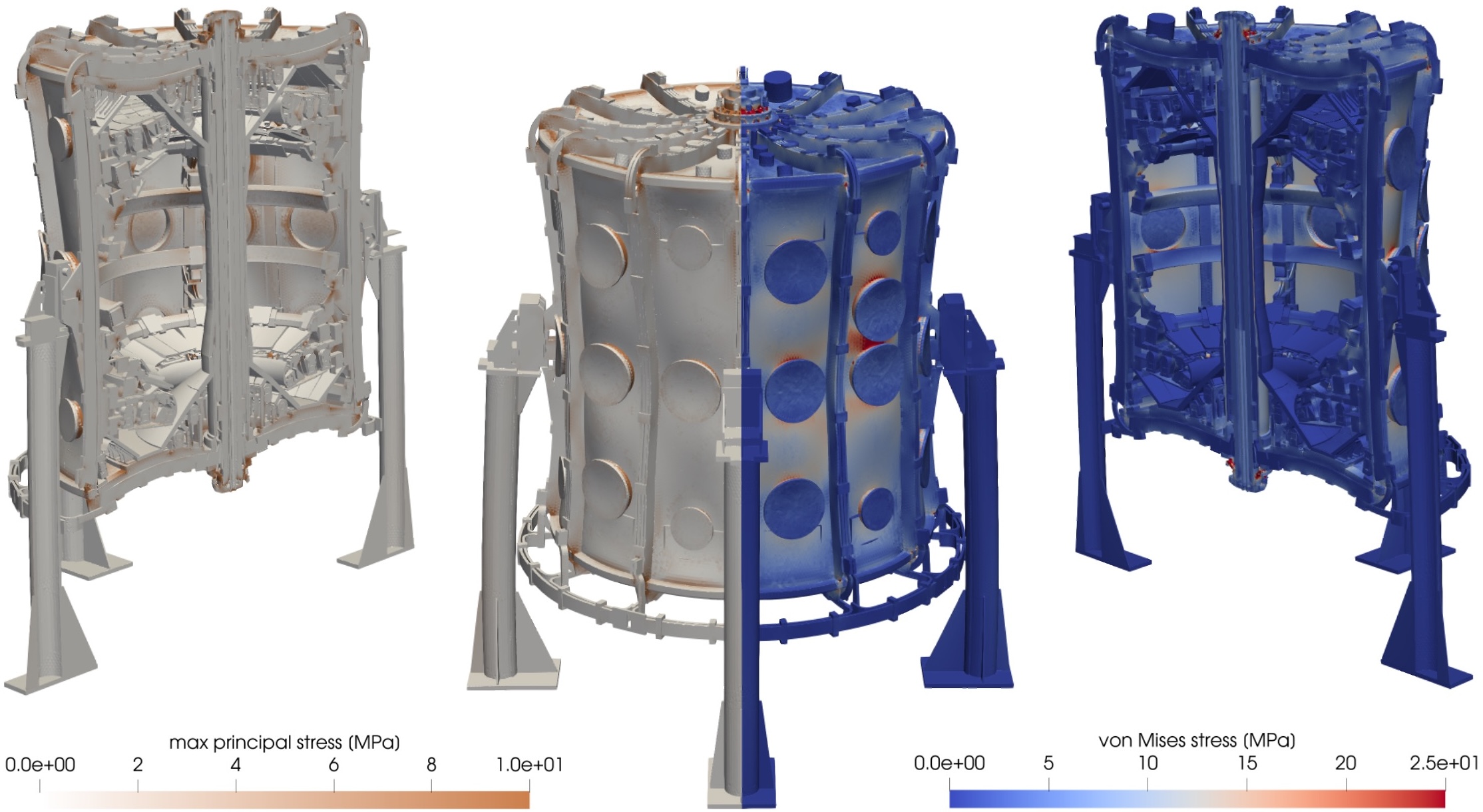}%
    \caption{Contour plot of stress in MAST-U under atmospheric pressure. The left-hand side of the panel displays the maximum principal stress while the right-hand side of the panel displays the von Mises stress. High stresses were found at the rim between the side and the top/bottom faces of the vacuum vessel, near the circular openings all around the side and where the central column prevents the inwards deformation of the top/bottom faces of the vacuum vessel. Deformation is magnified by a factor 1000 for demonstration.}
    \label{fig:atmos_stress} 
\end{minipage}
\vspace{2cm}
\begin{minipage}{1.\textwidth}
  \includegraphics[width=0.5\textwidth]{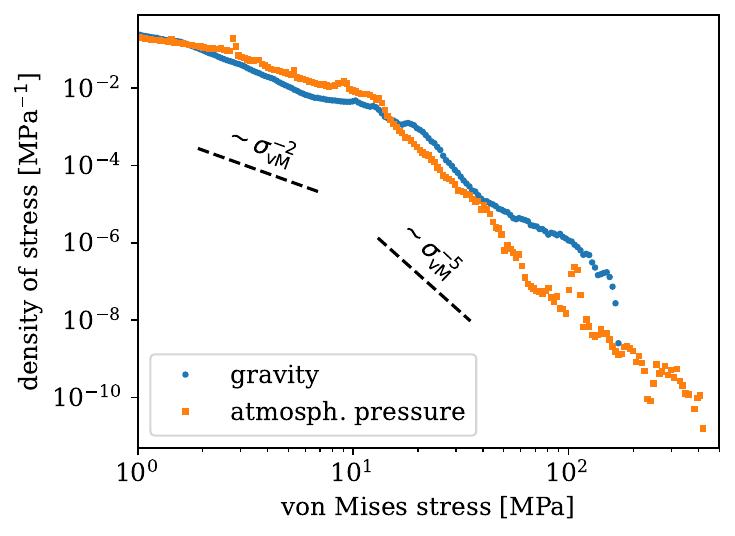} %
    \caption{Von Mises stress distribution (density of stress) calculated for the two simulations where either only gravity or only atmospheric pressure were used for loading the structure. In both cases, the distributions of points corresponding to a stress up to about \SI{10}{\mega\pascal} are close to a power law with exponent -2. Points below the \SI{1}{\mega\pascal} threshold, not shown, contributed a large fraction of the total integrated DOS (68\;\% and 56\;\% of the distribution).}
    \label{fig:DOS_MAST} 
\end{minipage}
\end{figure}

The peak values of von Mises stress, on the other hand, were higher and reached about \SI{400}{\mega\pascal}. Because of the bending of the vessel, side openings as well as the two circular rims experienced high stress, as well as the junction between the top/bottom surfaces and the central column. Maximum principal stress and von Mises stress are shown in Figure \ref{fig:atmos_stress} with an entire and two sectioned views of the reactor.

We now compare the action of gravity and atmospheric pressure on the full tokamak device using the concept of the density of stress (DOS), whose plot is shown in Figure \ref{fig:DOS_MAST}. In terms of volume, most of the elements experience low stress due to gravity, or atmospheric pressure. The density of von Mises stress for this low-stress region is well approximated by a power law with exponent close to -2. We found that 97\;\% of the elements under gravitational load and 96\;\% of the elements under atmospheric load were below the value of \SI{10}{\mega\pascal}. However, we also found a small fraction of elements that were reaching the stress values well above \SI{100}{\mega\pascal} in the very long tail of the distribution. The overall maximum was about \SI{170}{\mega\pascal} and \SI{420}{\mega\pascal} in the two simulations, respectively. On the basis of what is shown in Figure \ref{fig:DOS_MAST}, we conclude that, as expected, the stress induced by gravity and atmospheric pressure is \emph{globally} relatively low. This does not mean that these stresses should be discarded altogether since \emph{locally} some small volumes of the tokamak structure do experience substantial stress. Since adding gravitational and atmospheric pressure loads does not change the time to solution for any structural calculation, the above analysis shows that they ought to be considered alongside electromagnetic and thermal loads, and irradiation-induced stress.

\subsection{Frequency domain analysis}\label{sec:freq}
\begin{figure*}[t]
\centering
  \includegraphics[width=0.9\textwidth]{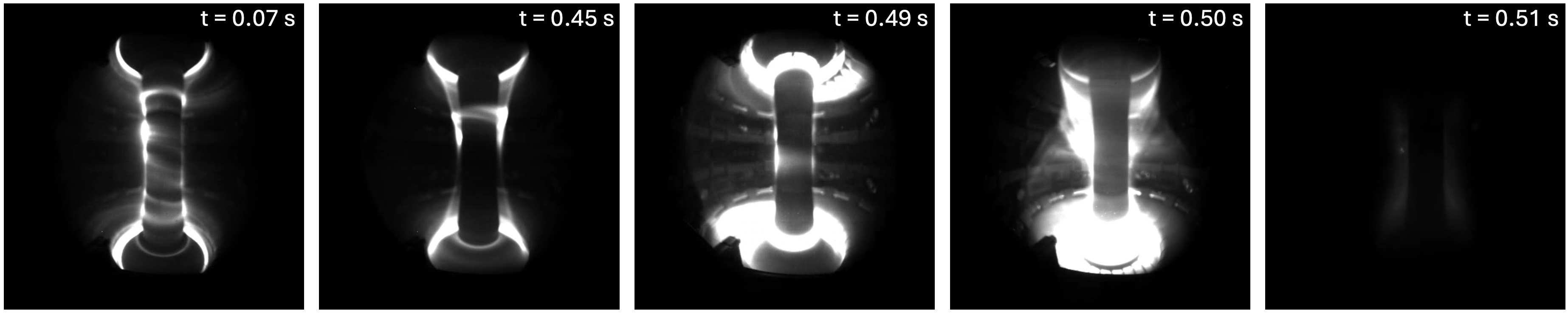}%

  \includegraphics[width=0.95\textwidth]{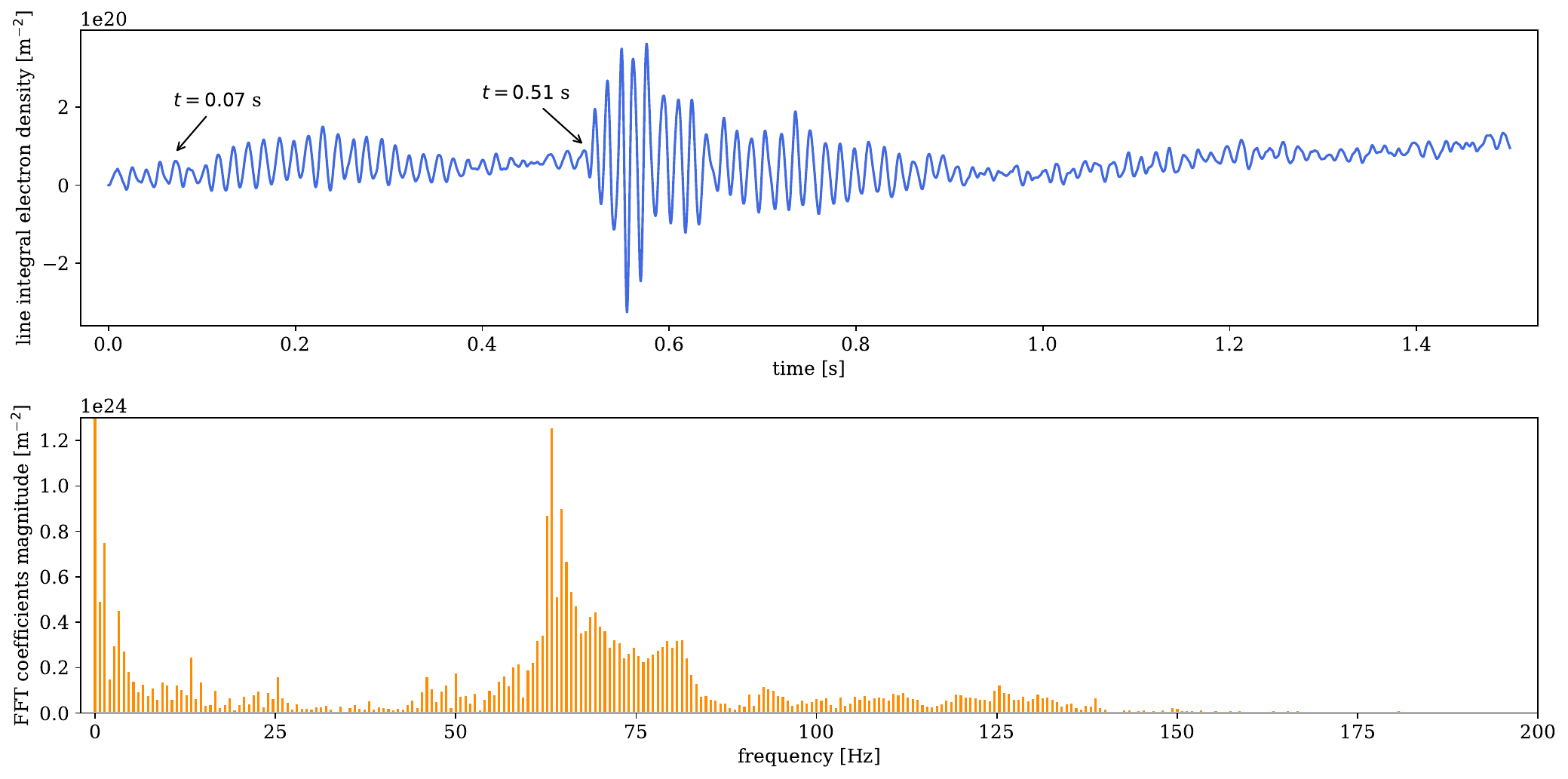}%
\caption{Experimental data from one of the pulses recorded on MAST-U. The row of images at the top shows snapshots from inside the tokamak (time is in seconds on the top right). The plot displays the signal from a laser interferometer mounted next to the reactor, top, and the fast Fourier transform of the signal, bottom. The plasma subsided at about $t=0.51$~s, after which a clear disruption in the signal can be observed.}
    \label{fig:exp_signal} 
\end{figure*}

A fusion reactor is a deformable structure that reacts dynamically to time-dependent loads. Research devices such as JET or MAST-U operate with single short pulses, and also ITER is expected to run in a pulsed mode, although the pulses are expected to be longer in duration. During a pulse, forces of electromagnetic origin generate time-dependent loads on the structure. Moreover, instabilities and abrupt termination of the plasma give rise to oscillatory motion of MAST-U itself and of the surrounding buildings and tokamak foundation. This is experimentally detected by an interferometer used for measuring electron density in the plasma, which clearly showed a mechanical signal consistent with an impact of plasma disruption. The signal, both in time and in frequency domains, is shown in Figure \ref{fig:exp_signal}. The measurement apparatus exploits laser interferometry, passing a HeNe laser beam through the plasma, and measuring a line integral plasma electron density (in units of m$^{-2}$), which is sensitive to mechanical vibrations \cite{Gornostaeva2010}. The measurements, alongside the necessity to develop a model of time-dependent electromagnetic forces, motivated a dynamic analysis in the frequency domain.

\begin{figure*}[h!]
  \includegraphics[width=0.75\textwidth]{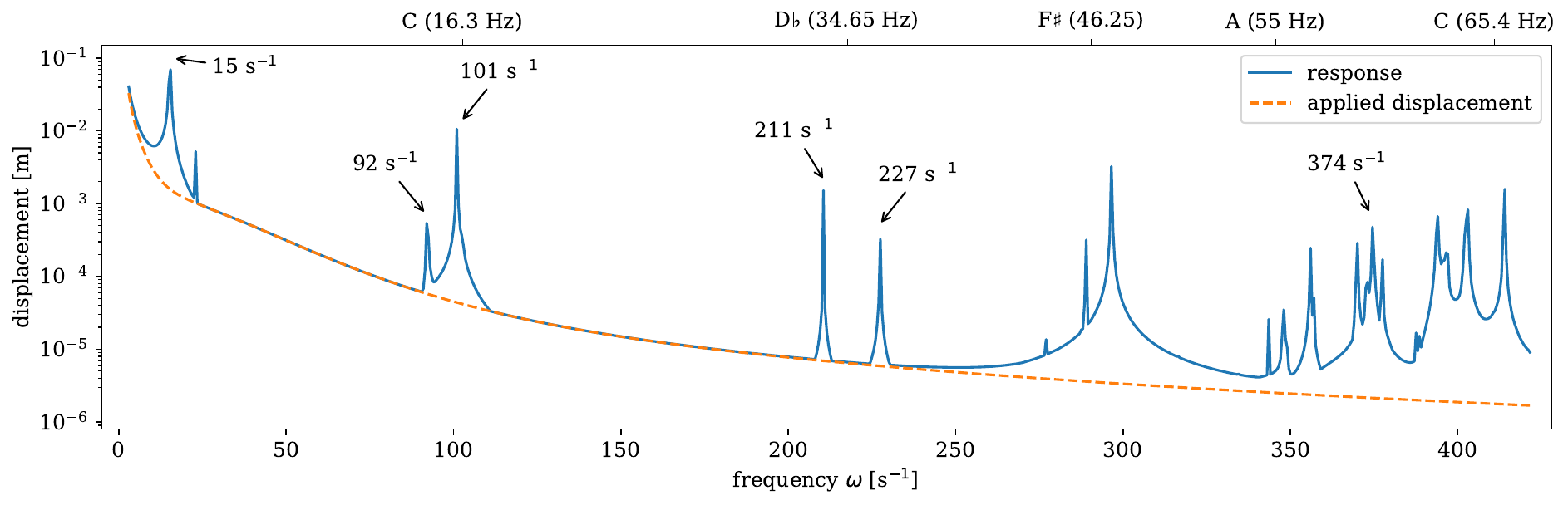} %
  
  \hspace{1cm}
  
  \includegraphics[width=.58\textwidth]{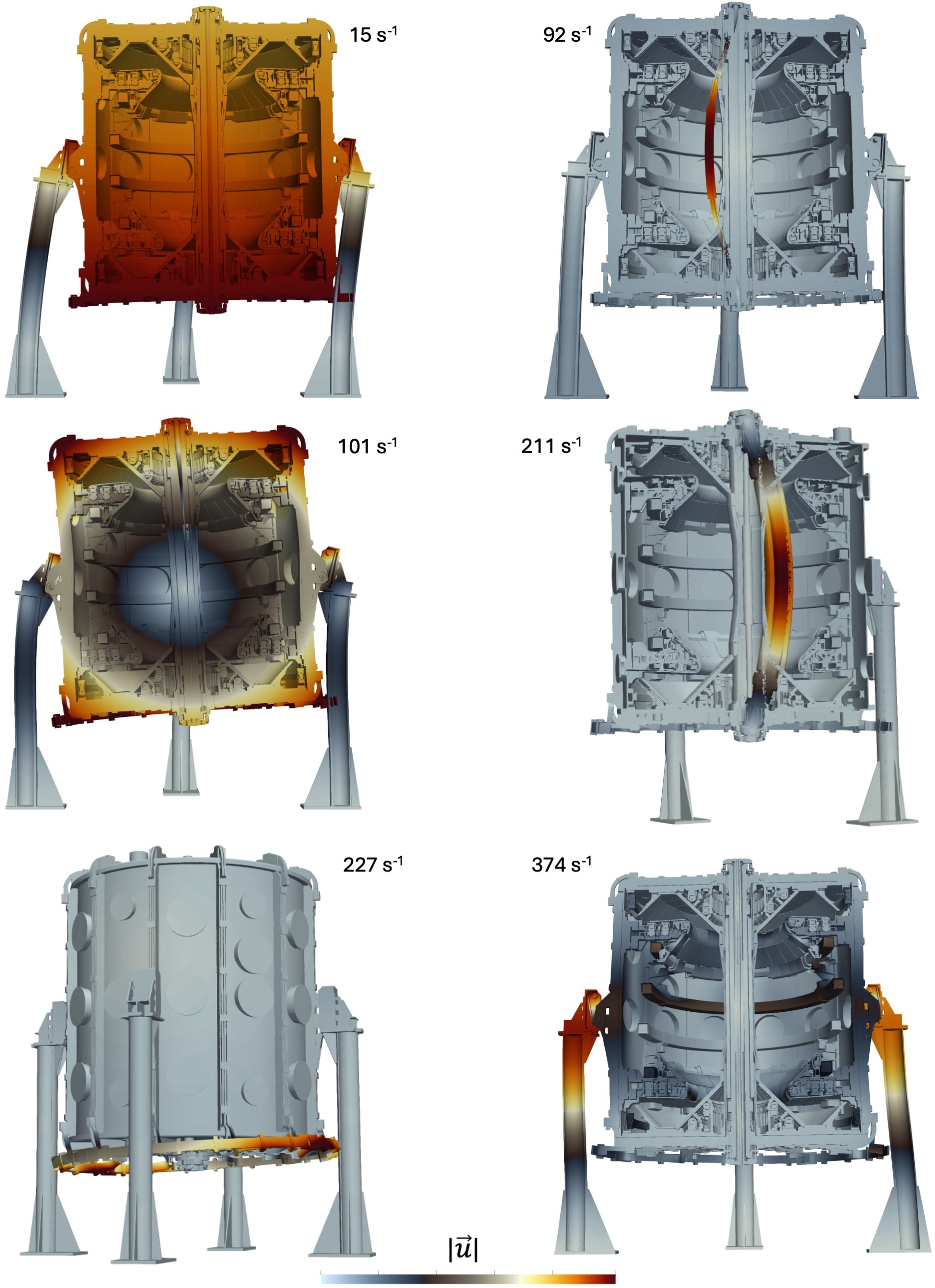} %
    \caption{Frequency response of the MAST-U tokamak structure to a harmonic horizontal displacement applied to its foundation. Solid line shows the maximum displacement magnitude anywhere in the structure given the displacement applied at the bottom of the legs, shown by the dashed line. Several resonances are clearly visible, involving the entire structure or some specific components, qualitatively associated with musical notes. Six of the resonances are illustrated  in the form of deformed structures, showing extremes of vibrational displacements, coloured by the magnitude of the displacement vector. The range of the colour bar and the scaling factor applied to make the displacements approximately comparable, vary between the figures. The use of scaling is necessary for visual purposes, to mitigate the considerable variation of scale in the graph shown at the top.}
    \label{fig:freq} 
\end{figure*}

To perform the analysis, we solved the three-dimensional Eq.~\eqref{eq:dynamic_frequency} using the full MAST-U finite element mesh and performed a linear elastic analysis in the small strain approximation. As a case study, we only applied displacement boundary conditions to the bottom surfaces of the four supporting legs. These were driven in an oscillatory motion along the horizontal $x$ direction, namely
\begin{equation}\label{eq:freq_BC}
    \begin{cases}
      x(t)=x_0(\omega)\sin(\omega t)\\
      y(t)=0 \\
      z(t)=0
    \end{cases}\,.
\end{equation}
where the $z$ axis is aligned with the central column, two of the legs are aligned with the $x$ and the other two with the $y$ direction. The amplitude of the forced displacement at the bottom of the machine, $x_0(\omega)$, was a function of frequency and is detailed in App.~\ref{app:x_0}. No other loads were applied.

The dynamic response of the tokamak structure was evaluated up to the frequency of 420 s$^{-1}$ in 0.5 s$^{-1}$ increments. In Figure \ref{fig:freq}, we plot the computed response spectrum of the full MAST-U structure. The plot shows the maximum displacement recorded at any of the nodes for a given frequency. We compare it to the displacement of the bottom of the legs, i.e. of the foundation of the device, which is imposed. For some frequencies, the entire structure moves by the same amount and the two curves overlap. In many cases, however, either the entire tokamak or some of its components resonate and the resulting displacement become greater than that of the foundation. 

The figure also shows detailed maps of the displacement field for six representative resonances. This gives another example of complexity associated with the full device: the greater the number of parts, the greater the number of potential resonances. If the structure were just a mass suspended on four springs \textemdash a good approximation for the first resonance \textemdash it would have displayed only a single resonance frequency. Adding a central column adds exitations similar to standing waves, such as those of a guitar string, and so on. Over the frequency range spanned by the analysis, we observed resonances of the entire vessel, of various concentric cylinders in the central column, of the poloidal and of the toroidal copper coils, and of the divertor tiles. In Appendix~\ref{app:res_model} we compare three of the resonances in the spectrum with simple mechanical analogies such as a mass suspended on springs or  vibrating elastic rods.

The analysis suggest that quantitative acoustic spectroscopy can be used as a detailed tool for monitoring the interaction of plasma with structural components of a tokamak device. This might be particularly beneficial for an operating fusion power plant where the use of conventional means for monitoring plasma and its interaction with the structure of the reactor is impeded by the effects of intense neutron and $\gamma$ radiation generated by the plasma and the first wall materials exposed to high energy neutrons \cite{Reali2023}.    

The acoustic eigenmodes of the device, once they have been established computationally and experimentally, can provide the basis for deconvoluting the result of plasma impacts on the various parts of the device, for example central column or the divertor. MAST-U provides a convenient test case for developing the methodology of acoustic spectroscopy since acoustic signals can be directly correlated with the visual control of plasma behaviour, also providing a suitable field for applying machine learning methods for correlating acoustic signals with visual monitoring of the plasma. The digital twin of the tokamak structure, similar to the model developed in this study, would then provide the predictive means for plasma control.


\section{Conclusions}

In this study, we developed an FEM model for a whole fusion tokamak device, in this instance the MAST-U spherical tokamak operating at the Culham Campus site of UKAEA. We started from a simplified geometric CAD model, initially created for neutron transport analyses, and not suitable for mechanical studies. By extensively modifying and extending the model, we have created a detailed finite element mesh describing the entire MAST-U device, containing all the features required for the mechanical analysis of the tokamak structure. The finite element mesh contains 127 million elements and involves structural parts made of stainless steel, copper and carbon. 

The model illustrates and highlights new fundamental issues encountered in massively parallel FEM simulations, which in our case involved close to 4,000 individual volumes and nearly 8,000 contact surfaces. The identification and correct representation of contact surfaces proved to be a critical time consuming issue that is not encountered in simulations involving a small number of components. Automatic contact detection algorithms proved instrumental, but offering room for improvement in error detection, to reduce the otherwise excessive need for human supervision. 

Using the new FEM model, we performed the first finite element simulations and explored mechanical deformations in a whole tokamak device. We investigated the effect of gravitational body forces and atmospheric pressure on the tokamak structure, finding that resulting stresses are overall low in magnitude but locally exceeding \SI{100}{\mega\pascal}. This shows that analyses of thermal, magnetic and radiation effects should include gravity and atmospheric pressure as well to provide reliable estimates for the distribution of stress and deformation in the structure. We also calculated the dynamical response of the tokamak subjecting its foundation to a horizontal oscillatory displacement of varying frequency, identifying a large number of resonances associated either with collective oscillation modes of the device, or with vibrations of its individual components. 
We also introduced the notion of the density of stress, to easily represent the state of stress in a complex engineering structure by a one-dimensional graph, with the physical meaning of the volume fraction of materials experiencing a given value of stress. This is especially useful in the limit where the complexity of the structure is significant and therefore its visual inspection has limitations. This study paves the way for further developments needed to account for the multi-physics and multi-scale nature of the problem of designing a fusion reactor, which also requires a treatment of electromagnetic forces as well as the use of accurate models for irradiation-induced eigenstrains and related changes in material properties.

We conclude by assessing the implications of the above analysis for modelling a full fusion
power plant. The FEM model for a MAST-U tokamak described above involves in excess of
$10^8$ tetrahedron elements. Taking the  800 m$^3$ plasma volume of ITER as a reference value,
and comparing this with the MAST-U plasma volume of 8 m$^3$, we can infer that a fusion
power-generating device of comparable complexity but a hundred times larger may require
a 10$^{10}$-element FEM model to simulate its structure at the same level of spatial resolution.

\noindent\textbf{Acknowledgements}\\
{\footnotesize
The authors gratefully acknowledge stimulating discussions with B. W. Spencer, R. Scannell, W. Morris, C. Ham, G. Aiello, A. Gray, and J. Buchanan. This work has been carried out within the framework of the EUROfusion Consortium, funded by the European Union via the Euratom Research and Training Programme (Grant Agreement No 101052200 — EUROfusion), and was partially supported by the Broader Approach Phase II agreement under the PA of IFERC2-T2PA02. This work was also funded by the UK Fusion Futures Programme and the UK EPSRC Energy Programme (grant number EP/W006839/1). To obtain further information on the data and models underlying the paper please contact PublicationsManager@ukaea.uk. Views and opinions expressed are however those of the authors only and do not necessarily reflect those of the European Union or the European Commission. Neither the European Union nor the European Commission can be held responsible for them. Numerical calculations, forming a part of this study, were carried out using resources provided by the Cambridge Service for Data Driven Discovery (CSD3).
}

\appendix
\section{Imprinting and Merging}\label{ImprintingAppendix}
\begin{figure}[h!]
    \begin{minipage}{.4\textwidth}
        \subfloat
        []
        {\label{fig:UnmeshedBoxes}\includegraphics[width=0.8\textwidth]{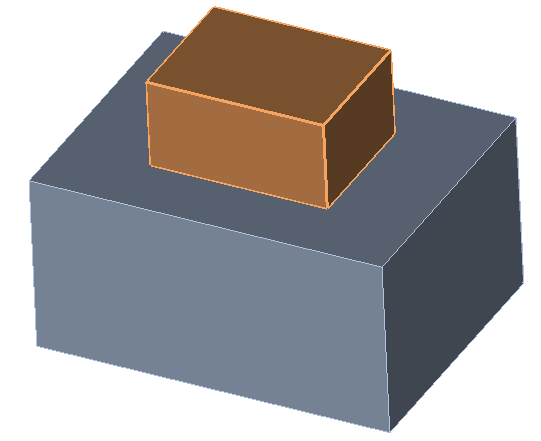}}
    \end{minipage}
    \begin{minipage}{.33\textwidth}
        \subfloat
        []
        {\label{fig:BoxNoImprint}\includegraphics[width=0.6\textwidth]{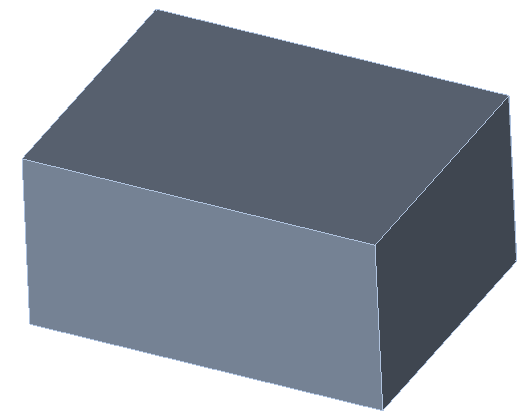}}
        
        \subfloat
        []
        {\label{fig:BoxImprint}\includegraphics[width=0.6\textwidth]{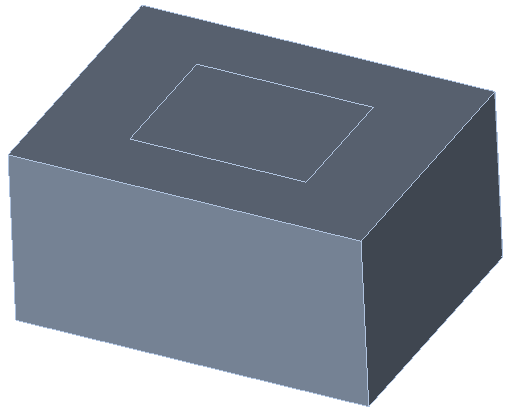}}
    \end{minipage}
    \caption{An example of how imprinting forces like topology using two cuboids, shown in a), that have two coincident surfaces. The curves from the smaller cuboid are projected onto the top surface of the larger cuboid, creating a new surface which has like topology with the smaller cuboid. The bottom cuboid pre- and post-imprint are shown in b) and c). The surfaces with like topology can now be merged.}
    \label{fig:Imprinting example}
\end{figure}

\begin{figure}[h!]
    \begin{minipage}{.4\textwidth}
        \subfloat
        []
        {\label{fig:MeshedBoxes}\includegraphics[width=0.8\textwidth]{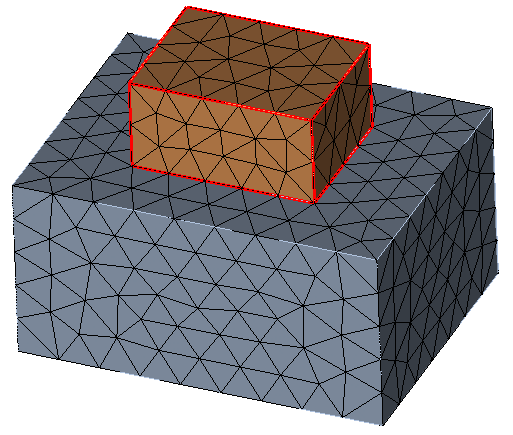}}
    \end{minipage}
    \begin{minipage}{.33\textwidth}
        \subfloat
        []
        {\label{fig:MeshedBoxesNoImprint}\includegraphics[width=0.6\textwidth]{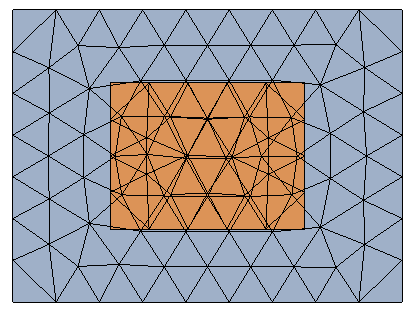}}
        \subfloat
        []
        {\label{fig:MeshedBoxesImprint}\includegraphics[width=0.6\textwidth]{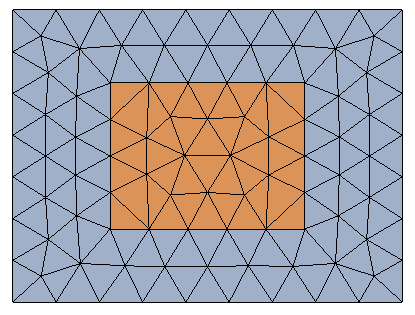}}
    \end{minipage}
    \caption{a) The same geometry as Figure \ref{fig:Imprinting example} is used to show how imprinting forces conformal mesh elements across component boundaries. The meshed co-planar surface shared between the two cuboids, without imprinting, is shown in b). The meshes from each cuboid do not conform to one another as they are not imprinted and merged. The meshed co-planar surface shared between the two cuboids after imprinting is shown in c). The resultant mesh is desirable as the finite elements are conformal at the boundary.}
    \label{fig:OverlappingMeshes}
\end{figure}

Figure \ref{fig:Imprinting example} demonstrates the process of imprinting CAD geometry. The initial geometry in this example is comprised of two cuboids with the smaller stacked on top of the larger, hence having a co-planar surface. Prior to imprinting the two cuboids lack shared topology. Figure \ref{fig:BoxImprint} shows that after imprinting the curves from the smaller cuboid on the co-planar surface  are projected on to the top surface of the larger cuboid. The projection of these curves results in a new surface on the larger cuboid which shares topology with the smaller cuboid. The surfaces with like topology can now be merged together.  
 The effect of imprinting on the resulting mesh is shown in Figure \ref{fig:OverlappingMeshes}. Prior to imprinting the finite elements on the cuboid meshes do not conform to each other, as shown in Figure \ref{fig:MeshedBoxesNoImprint}. After imprinting a desirable mesh is achieved with conformal finite elements across the component boundaries.

\section{DOS for a cantilever beam}\label{app:beam}
Consider a slender beam of length $L$ and rectangular cross section. The origin of the coordinate system is at the clamping position in the centre of the cross section; $z$ is along the axis of the beam and gravity acts along negative $y$ direction. The beam has thickness $t$ along $y$ and $w$ along $x$. Moreover, there is a concentrated force at the free end corresponding to the total weight of the beam. If the length is much longer than the thickness in the two orthogonal directions and Euler-Bernoulli beam theory applies \cite{Timoshenko1983}, all stress components are negligible with respect to the bending stress $\sigma_{zz}$ whose expression is
\begin{equation}
    \sigma_{zz}=\frac{6\rho g y}{t^2}(3L^2-4Lz+z^2).
\end{equation}
The stress has symmetric maximum compression and tension at the clamping position and top and bottom surfaces ($z=0$, $y=\pm t/2$), of magnitude $\sigma_{\text{max}}=9\rho g L^2/t$.

To calculate the DOS, we start from definition \eqref{eq:DOS_definition} and write
\begin{equation}
    D(\sigma)=\frac{1}{Lwt}\int_0^w\int_{-t/2}^{t/2}\int_0^L \text{d}x\text{d}y\text{d}z \;\; \delta\left(\sigma-\frac{6\rho g y}{t^2}(3L^2-4Lz+z^2)\right).
\end{equation}
The integrand does not depend on $x$ therefore the integral over $\text{d}x$ cancels the $w^{-1}$ factor. We then make the substitution $u=\sigma-6\rho g y(3L^2-4Lz+z^2)/t^2$, change the bounds of the integration over $y$ accordingly and then extend the range of integration to $(-\infty, \infty)$ by introducing two step functions. The Dirac delta function can has us pick the integrand at $u=0$ and we are left with
\begin{equation} \label{eq:DOS_H}
    D(\sigma)=\frac{1}{2\sigma_{\text{max}}}\frac{1}{L}\int_0^L\frac{\text{d}z}{1-\frac{4z}{3L}+\frac{z^3}{3L^2}}H\left(\sigma_{\text{max}}\left(1-\frac{4z}{3L}+\frac{z^3}{3L^2}\right)-\sigma\right)H\left(\sigma_{\text{max}}\left(1-\frac{4z}{3L}+\frac{z^3}{3L^2}\right)+\sigma\right).
\end{equation}
Function $1-\frac{4z}{3L}+\frac{z^3}{3L^2}$ monotonously decreases from 1 to 0 as z varies from 0 to $L$. Hence, for $\sigma<-\sigma_{\text{max}}$ and $\sigma>\sigma_{\text{max}}$ the integrand always vanishes. For $0<\sigma<\sigma_{\text{max}}$, the second step function is always 1. To find where the first step function affects the integral we start by finding that the roots of its argument are $2L\pm L\sqrt{1+2\sigma/\sigma_{\text{max}}}$. Since $0\leq z\leq L$ we keep only the one denoted as $a=2L-L\sqrt{1+2\sigma/\sigma_{\text{max}}}$. This allows us to write the integral \eqref{eq:DOS_H} and solve it:
\begin{equation}
    D(\sigma)=\frac{1}{2L\sigma_{\text{max}}}\int_0^a\frac{\text{d}z}{1-\frac{4z}{3L}+\frac{z^3}{3L^2}}=\frac{3}{4\sigma_{\text{max}}}\log\left(\frac{a-3L}{3a-3L}\right).
\end{equation}
We repeat the same calculation in an analogous way for $-\sigma_{\text{max}}<\sigma<0$. Substituting the definition of $a$ and manipulating the argument of the logarithm we arrive at the final result of eq.\eqref{eq:DOSbeam}.

\section{Forced displacement amplitude}\label{app:x_0}
\begin{figure}[t]
  \includegraphics[width=0.45\columnwidth]{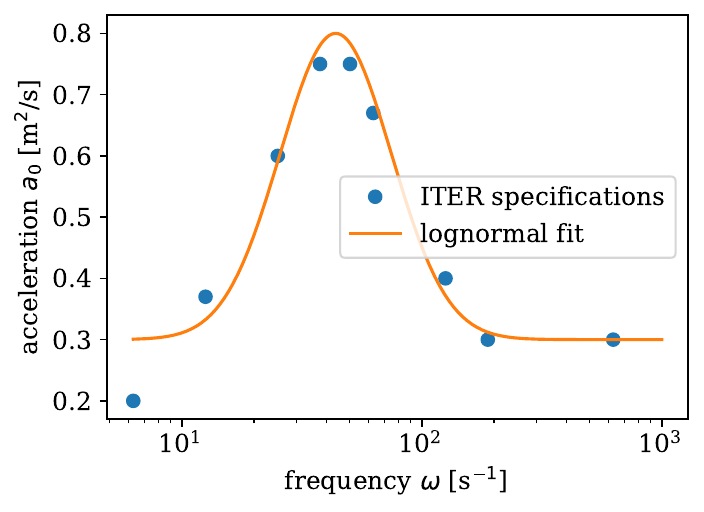} %
    \caption{Amplitude of ground acceleration assuming 5\% damping: discrete data from Figure 3 of \cite{Bachmann2017} and lognormal fitting function $a_0=\frac{1}{2}\rm{exp}[-1.75\rm{log}(\omega-3.78)]^2+0.3$.}
    \label{fig:a_0} 
\end{figure}

In the frequency domain analysis of Sec.~\ref{sec:freq} the amplitude of the horizontal displacement $x_0$ in Eq.~\eqref{eq:freq_BC} depends on frequency. To calculate $x_0(\omega)$ we used the following approach. First, we take the ITER specifications for the seismic loads from Ref. \cite{Bachmann2017}, where the amplitude $a_0$ of the ground \emph{acceleration} is given for different values of damping. Here, we chose the intermediate 5\% value. For mathematical simplicity, we fitted a log-normal distribution to the data, noting that the lowest frequency that we considered in the analysis is \SI{2}{\hertz}. The data and the fitted curve are shown in Figure \ref{fig:a_0}. Second, we infer the displacement amplitude from the acceleration amplitude by integrating twice
\begin{equation}
    \ddot{x}(t)=a_0(\omega)\sin(\omega t)
\end{equation}
to obtain
\begin{equation}
    x(t)=-\frac{a_0(\omega)}{\omega^2}\sin(\omega t)
\end{equation}
We note first that seismic loads are hardly relevant to conditions in England, where MAST-U is located; they are used here for demonstration. Second, the analysis is linear in the applied displacement, hence the magnitude of the imposed displacement is somewhat immaterial as the response spectrum to a different displacement applied to the bottom supports can be obtained by simple proportionality.  

\section{Simple models of some resonances}\label{app:res_model}
It is possible to compare some of the resonances with very simple mechanical models. At the first resonance, centred at 15 s$^{-1}$, MAST-U can be modelled as a concentrated mass $m$ suspended on four springs of stiffness $k$ reacting in parallel. The resonant frequency of this system is $\omega=\sqrt{4k/m}$. The mass is the entire tokamak mass minus that of the four legs, i.e. $9.2\cdot10^4$ kg. The four legs are modelled as hollow cylinders, which have a combined bending stiffness of $3EJ/L^3$, where $E$ is the Young's modulus, $J=\frac{\pi}{4}(R_o^4-R_i^4)$ is the inertia of the cross-section and $L$ is the bending length. The legs have outer radius $R_o=\SI{178}{\milli\meter}$ and inner radius $R_i=\SI{159}{\milli\meter}$. Their total length is about \SI{4}{\meter}, but the stiffening ribs, reaching approximately \SI{1}{\meter} from the ground, mean that not all of this length can freely bend. If we assume a bending length $L=\SI{3}{\meter}$ we find a lower bound of 10.9s$^{-1}$, whereas assuming $L=\SI{4}{\meter}$ gives the upper bound of 17.1 s$^{-1}$. The calculated resonance of 15 s$^{-1}$ is inside this range. 
\begin{figure}[t]
  \includegraphics[width=0.9\columnwidth]{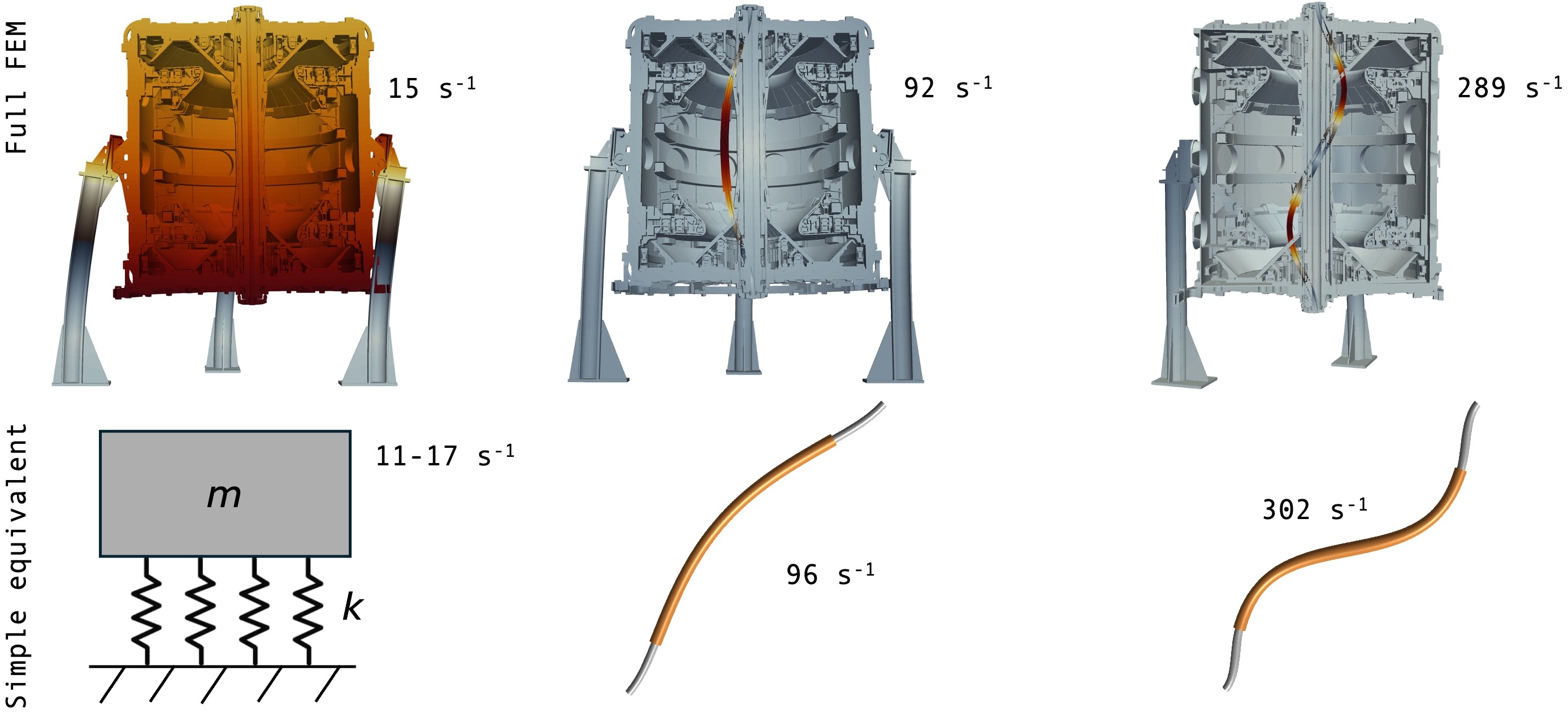} %
    \caption{Three of the resonances in the acoustic response spectrum of MAST-U are compared with simplified mechanical models. The resonance at 15\;s$^{-1}$ is compatible with the resonance of a mass suspended on 4 springs. The first resonances of the innermost steel rod that suspends a copper coil are compared to the eigenmodes of the same component, if taken in isolation as a doubly-clamped beam, extracted through a separate FEM analysis.}
    \label{fig:res_models} 
\end{figure}
The second resonance of the spectrum, at 92 s$^{-1}$, is the first standing wave of the innermost steel rod of the central column, which supports a copper coil that is about one third of the length. The second standing wave is found at 289~s$^{-1}$. Also here we can propose a simple mechanical analogy, which is approximating the components as co-axial cylinders constrained at both ends. For example, the inner component is a steel hollow rod ($R_o=\SI{51}{\milli\meter}$, $R_i=\SI{28}{\milli\meter}$, $L=\SI{5.3}{\meter}$) that is clamped at both ends, i.e. both displacements and curvature at the ends are forced to be zero. If the hollow rod were in isolation its $n^{\rm{th}}$ standing wave would be given by $\omega_n=((n+1/2)\pi/L)^2\sqrt{EJ/(\rho A)}$, where $\rho$ is the density and $A$ the cross-section area. The rod however is supporting a copper coil (of width \SI{260}{\milli\meter}, extending between \SI{800}{\milli\meter} from both ends) that increases both the bending stiffness and the mass. Ignoring the copper coil would place first and second resonances at 119~s$^{-1}$ and 331~s$^{-1}$. The copper adds mass concentrated in the central section of the rod, but it is about half as stiff as steel, hence these two values are overestimating the expected ones. We therefore extracted just the rod and the copper coil, clamped the steel rod at both ends and calculated the eigenmodes of the system using the Abaqus programme. The first standing wave was found at 96.1\;s$^{-1}$ while the second was found at 302.1\;s$^{-1}$, slightly lower than the analytical estimate as expected and within about 4\% of the corresponding peaks in the MAST-U spectrum. A summary of the analysis is presented in Figure \ref{fig:res_models}.
\bibliography{reference}

\end{document}